\documentclass{article}
\usepackage{float}
\usepackage{float}
\usepackage{float}
\usepackage{colortbl}
\usepackage{array}
\usepackage{tabularx}
\usepackage{abstract}
\usepackage{float}
\usepackage{abstract}
\usepackage{amsmath}
\usepackage{authblk}
\usepackage{booktabs}
\usepackage[table]{xcolor}

\newcommand{\undertitle}[1]{\def\@undertitle{}}
\makeatletter

\usepackage{arxiv}
\usepackage{authblk}

\usepackage[utf8]{inputenc} 
\usepackage[T1]{fontenc}    
\usepackage{hyperref}       
\usepackage{url}            
\usepackage{booktabs}       
\usepackage{amsmath}        
\usepackage{amsfonts}       
\usepackage{nicefrac}       
\usepackage{microtype}      
\usepackage{lipsum}		    
\usepackage{graphicx}
\usepackage{doi}
\usepackage{xcolor}
\usepackage[table]{xcolor}
\usepackage{multirow}
\usepackage{subcaption}
\usepackage{array}
\usepackage[round]{natbib}
\usepackage{graphicx} 
\usepackage[skip=0.333\baselineskip]{caption}
\usepackage{subcaption}
\usepackage[font=small,skip=0pt]{caption} 

\begin{document}
\title{What Can 240,000 New Credit Transactions Tell Us About the Impact of NGEU Funds?}\thanks{The authors thank, without implication, Eduardo Aguilar, Tamara de la Mata, and Iñigo Portillo from the Directorate General of the Spanish Ministry of Economy. We also extend our gratitude to Enrique Moral-Benito and the seminar participants at the Bank of Spain, as well as Vasco Carvalho, Stephen Hansen, Sevi Rodriguez-Mora, Raúl Pérez, Miguel Cardoso, Pep Ruiz and Virginia Pou for their valuable comments.}


\author{
    {\hspace{1mm}\textbf{Alvaro Ortiz}}\\
	BBVA Research \& CRIW \\ 
      \vspace{2mm}
    \texttt{alvaro.ortiz@bbva.com} \\
  
	{\hspace{1mm}\textbf{Tomasa Rodrigo}} \\
	BBVA Research \\
    \vspace{2mm}
	\texttt{tomasa.rodrigo@bbva.com} \\
    
    {\hspace{1mm}\textbf{David Sarasa}} \\
	BBVA Research \\
    \vspace{2mm}
	\texttt{David.Sarasa@bbva.com} \\

    {\hspace{1mm}\textbf{Sirenia Vazquez}} \\
	BBVA Research \\
    \vspace{2mm}
	\texttt{sirenia.vazquez@bbva.com} \\
}
 
\maketitle
\vspace{-1\baselineskip}

\begin{abstract}Using a panel data local projections model and controlling for firm characteristics, procurement bid attributes, and macroeconomic conditions, the study estimates the dynamic effects of procurement awards on new lending, a more precise measure than the change in the stock of credit. The analysis further examines heterogeneity in credit responses based on firm size, industry, credit maturity, and value chain position of the firms. The empirical evidence confirms that public procurement awards significantly increase new lending, with NGEU-funded contracts generating stronger credit expansion than traditional procurement during the recent period. The results show that the impact of NGEU procurement programs aligns closely with historical procurement impacts, with differences driven mainly by lower utilization rates. Moreover, integrating high-frequency financial data with procurement records highlights the potential of Big Data in refining public policy design.
\end{abstract}


\newpage

\section{Introduction} 
\label{sec:Introduction}

Public procurement has long been recognized as a key driver of economic activity, facilitating firm growth, financial stability, and investment. As governments allocate a significant portion of GDP to procurement - more than 14\% in the EU \citep{EC} - it serves as a crucial policy tool to support businesses, particularly small and medium companies (SMEs) that rely on government contracts for stability and expansion. By financing infrastructure projects, industrial modernization, and digitalization efforts, procurement injects liquidity into the private sector, influencing firm credit behavior and overall economic resilience. However, the extent to which procurement awards translate into increased credit availability remains an open question, especially with respect to new lending operations rather than changes in credit stock.

The Next Generation EU (NGEU) program has placed public procurement in even sharper focus as a mechanism for economic recovery and transformation. Launched in response to the COVID-19 crisis, NGEU represents the largest economic stimulus package in European history, mobilizing €750 billion to support digital and green transitions, enhance financial stability, and stimulate investment\footnote{From which €163 billion were allocated to Spain (€80 billion in grants and €83 billion in loans) \citep{NGEUSpain}. By March 2024, and in accordance with the calculations provided by the official government tracker of the call and execution of NGEU funds in Spain \citep{ELISA}, approximately €80 billion have been called, while the awarded amount totals nearly €49 billion, which represents a resolution rate of 60\%.}. Given its scale, procurement serves as a primary channel for deploying NGEU funds, particularly in hard-hit economies. Assessing whether firms that secure NGEU-funded procurement contracts experience a stronger credit response compared to traditional procurement awards is crucial to evaluating the effectiveness of the program in fostering financial recovery and firm growth.

We examine how procurement awards impact firm-level new lending operations in Spain and whether the NGEU framework amplifies this effect. Using a panel data Local Projections model, we analyze a novel dataset merging public procurement records with high-granularity new lending data from BBVA, a major Spanish bank. This unique integration captures firm-level borrowing responses to procurement contracts in high detail.

By assessing sector, firm size, credit maturity, and value chain position, we isolate the effects of NGEU versus traditional procurement to determine whether targeted European recovery funds enhance firm liquidity. These findings inform procurement policy and fiscal interventions to optimize credit expansion and economic growth.

We contribute to the literature by leveraging a unique large-scale dataset that links public procurement awards, particularly NGEU-funded contracts, with firm-level new credit data from BBVA.\footnote{BBVA is one of Spain’s largest lenders, ranking second by total assets.} Unlike previous studies that analyze total outstanding credit, we focus on new lending operations, offering a more precise measure of how procurement contracts influence firm financing. This distinction is crucial, as changes in credit stock can be affected by amortization and other financial activities, potentially obscuring the direct impact of procurement on liquidity. Furthermore, by examining firm-level heterogeneity, we identify differential credit responses based on firm size, industry, and value chain position, providing new insights into how procurement-induced credit expansion varies across firms

This paper finds that public procurement awards lead to a significant increase in firm-level new lending, with a cumulative increase of 0.75\% in new credit operations within one year during the period from August 2019 to July 2024. However, the impact of NGEU programs has been substantially higher, reaching nearly 3.0

When adjusting for the amortization effect to ensure comparability with standard credit stock changes, the resulting one-year dynamic elasticity of credit stock is approximately 1.5 percentage points for overall procurement programs during the same period controlling for procurement attributes. The effect becomes evident six months after contract awards and persists throughout the first year. However, the elasticity for NGEU-funded procurement is significantly higher, close to 5\%, and remains within the range of historical estimates for Spain (6\%), as reported by \cite{diGiovanni2022}. This suggests that the impact of NGEU procurement on credit expansion aligns with the historical trend.\footnote{In terms of the overall budget distribution within the Next Generation EU (NGEU) framework, approximately 45\% of the total funds are allocated via public procurement instruments, while the remaining 55\% are delivered through grants. In this paper, we concentrate exclusively on examining the effects of the procurement channel. Furthermore, our estimations corresponds to the procurement part of the NGEU program for Spain.}

In sum, NGEU-funded contracts have generated a significantly stronger credit response compared to no-NGEU procurement during August 2019 to July 2024. Moreover, the estimated elasticity remains consistent with historical patterns.

There is significant heterogeneity in the impact of procurement-induced credit expansion across firm size, industry, and value chain position. Smaller firms and those in government-dependent sectors, such as construction and manufacturing, experience the largest credit boosts, while firms in less government-dependent industries exhibit more muted responses. Furthermore, short-term credit expands more than long-term credit, suggesting that firms primarily use procurement-induced financing for working capital rather than long-term investment.

Unlike prior studies, we find no anticipation effects, meaning that firms adjust borrowing only after securing contracts. This heterogeneity highlights the role of firm characteristics in shaping the effectiveness of procurement in firm financing and the broader macroeconomic impact of NGEU funds in supporting economic recovery.

We also identify differences in credit responses to procurement awards based on a firm's position in the value chain. Applying the seamless position analysis developed \cite{antras2012}, we find that latest-stage downstream firms, which are closer to final demand, experience a more immediate credit expansion, whereas early upstream firms experience a delayed yet more sustained increase in credit availability. 

Our findings suggest that the effects of procurement would gradually propagating through the production network. However, unlike the pronounced disparities observed in the transmission of monetary policy in Spain \cite{buda2023}, these differences are statistically significant only at the 68\% confidence level and are most evident at the extremes of the value chain, particularly when comparing upper early upstream and lower latest-stage downstream segments. Further, the observed variations in credit responses result from idiosyncratic procurement shocks rather than a monetary common policy shock.

This paper is structured as follows. Section 2 describes the data sources and the construction of our novel Big Data set. Section 3 outlines the empirical strategy, explaining the panel data Local Projections model used to estimate the dynamic effects of procurement awards on new lending while controlling for firm and procurement characteristics as well as macroeconomic conditions. Section 4 presents the main results, detailing the aggregate effects of procurement on firm credit, the differential impact of NGEU vs. no-NGEU contracts, and the heterogeneous responses by firm size, sector, credit maturity, and value chain position. Finally, Section 5 concludes by summarizing the findings and suggesting avenues for future research on procurement-driven credit expansion.

\textbf{Literature review} 

This paper contributes to the broader literature on the effects of government spending on economic growth and the size of fiscal multipliers. Studies using both macro and microdata provide mixed evidence on the relationship between fiscal policy and output growth (\cite{gali2007,ramey2011,gabriel2024credit,briganti2024effects}). A key dimension of this debate concerns the role of financial frictions in shaping the response to fiscal shocks. Research has shown that the impact of government spending is stronger and more persistent when credit constraints are binding, either at the macroeconomic level during financial crises \cite{ferraz2015} or in specific regions and sectors characterized by limited access to credit (\cite{aghion2014,juarros2020}).

Within this field, studies have examined the credit channel of fiscal stimulus, particularly in the context of targeted government interventions such as the Troubled Asset Relief Program (TARP) \cite{duchin2014} and the 2004 American Jobs Creation Act \cite{bird2022}. 
The empirical evidence presented in this paper regarding the treatment effects of public procurement awards on firm performance aligns with recent scholarly literature on the impact of winning procurement contracts on firm dynamics. Focused on firm growth enhancement, \cite{ferraz2015} and \cite{lee2021} suggest that firms that secure procurement contracts exhibit greater growth compared to their competitors, based on data from Brazil and Korea, respectively. Furthermore, \cite{hebous2021can} document a positive correlation between winning a procurement contract and firm investment in the U.S., although this effect diminishes when examining firms that are less likely to face financial constraints.

Relative to the existing empirical literature on public procurement in Europe, \citet{diGiovanni2022} and \citet{gabriel2024credit} show that public procurement can act as a financial catalyst for firms by enhancing their access to credit. \citet{diGiovanni2022}, using Spanish firm and public procurement data from 2000 to 2013, demonstrates that procurement contracts can serve as collateral, helping firms—especially small and financially constrained ones—overcome borrowing limits and expand their operations.

Compared to their findings, our results show a lower but comparable elasticity for all procurement programs. However, we find a slightly lower (5\%) consistent elasticity for NGEU-funded procurement bids relative to the historical elasticity estimate of approximately 6\% reported by \citet{diGiovanni2022}.

These findings reinforce the idea that procurement can enhance financial stability and firm growth, particularly in liquidity-constrained environments. Our results are in line with \citet{gabriel2024credit}, who explores the credit channel of public procurement in Portugal using a novel dataset covering public procurement and firm characteristics from 2009 to 2019. His study provides further empirical evidence that winning a public contract increases firm credit availability and reduces borrowing costs, reinforcing the role of public procurement as a credit supply shock.

While extensive research has explored the impact of public procurement on firm dynamics, empirical evidence on the economic effects of Next Generation EU (NGEU) funds remains limited. In Spain, \citet{aguilar2023primera} provide one of the first analyses of firms receiving NGEU funds, finding that these firms are generally larger, more productive, and have better access to bank financing compared to those awarded traditional procurement contracts. At the macroeconomic level, \citet{ecb2020} assess the broader implementation of the NGEU program and highlight that its expected benefits have been weakened by lower utilization rates rather than efficiency problems.

Finally, our paper contributes to the rapidly expanding literature on high-frequency and high-granularity databases and indicators, driven by the need for timely policy decisions in fast-changing environments such as the COVID-19 pandemic. Examples include weekly indicators (\cite{eraslan2021,baumeister2022tracking} \cite{Lewis2020}) and daily indicators (\cite{diebold2022,rua2020}). Alongside these, a growing number of studies utilize transaction-based data to capture real-time shifts in economic activity or conduct highly detailed distributional analyses using financial transaction data  (\cite{andersen2022,Buda2022,Chetty2020,ganong2019,barlas2024})


\section{The Data: A Public-Private Big Data Base of Public Procurement and New Lending Operations}
\label{sec:thedata}

In this section, we detail the construction of our final database, which is designed to analyze the impact of public procurement awards on new lending operations by sector of activity. Specifically, we distinguish between procurement contracts financed by the Next Generation EU (NGEU) program and other public procurement bids. 

To achieve this, we integrate multiple public and private data sources in a multi-step process, ensuring a comprehensive and high-quality dataset. To the best of our knowledge, this is the first dataset that provides daily data on new corporate lending operations, combined with firm-level public procurement awards.

Our data construction follows three key steps:

\begin{enumerate}
    \item \textbf{Public Procurement Data}

We incorporate data on public procurement tenders from the Ministry of Finance of Spain, obtained from its official public procurement portal\footnote{Official data can be found on the public portal of the Ministry of Finance e in the following link https://www.hacienda.gob.es/esES/GobiernoAbierto/Datos\%20Abiertos/Paginas/licitacionesplataformacontratacion.aspx. This dataset distinguishes between procurement projects funded by the NGEU program and non-NGEU public contracts at the firm level. The final procurement dataset consists of 381,000 observations at a daily frequency, covering approximately 100,000 firms that participated in government procurement between August 2019 and July 2024. From these firms, we obtained their sector of activity according to the industry standard classification system used within the European Union (NACE code 2-digits); the number of awards (NGEU and non-NGEU), as well as the amount, the publication date, execution period, and the authority or public entity in charge of managing and publishing the tenders.}

    \item \textbf{New Lending Operations} 

To capture firm-level lending activity, we use proprietary data on new corporate lending operations from BBVA, one of Spain’s largest banks. BBVA holds a market share in corporate lending around 14\%. The dataset includes 5,090,000 credit transactions covering 318,000 firms, with daily observations over the same period (August 2019 – July 2024). This allows us to assess the responsiveness of corporate borrowing to procurement contracts. From this, we are using the credit amount, the weighted interest rate by credit amount and well as the number of credits. We also disentangle between long-term credit, sum of short-term credit and ICO credits, that are loans provided by the Instituto de Crédito Oficial (ICO), which is a Spanish public financial institution, designed to support businesses and economic activities by offering favorable financing conditions.

  \item \textbf{Firm Characteristics}
 
To enrich our dataset with firm-level characteristics, we merge it with SABI, a comprehensive financial and business database covering over 2.9 million Spanish firms. SABI provides key firm attributes, including industry sector, age, turnover and employment, which are crucial control variables for our analysis of how NGEU funding influences corporate lending. From this external database, we are considering the company reported revenue, the number of employees, their turnover, net capital, financial rating and birth year.

 \end{enumerate}

Finally, to build the final dataset, we match firms across the different datasets using their tax identification code to have their credit performance, if they have been funded by the NGEU program, no-NGEU public contracts or not, as well as the socioeconomic features of the firms. Given the nature of our analysis, we aggregate data at a monthly frequency. Table \ref{tab:Tabla1} summarizes the results. The final dataset consists of 2,062 firms receiving NGEU funding (92\% of which also received no-NGEU contracts) and 17,282 firms engaged in no-NGEU public procurement. Therefore, we work with 1,045,980 total observations, including 239,154 credit transactions and 119,322 public procurement awards.

This uniquely structured dataset allows us to investigate how firms' borrowing behavior responds to procurement contracts, while differentiating the role of NGEU funds compared to other public procurement programs.

\begin{table}[h]
    \centering
    \caption{\textbf{Descriptive statistics: the Database}}
    \vspace{0.3cm}
    \label{tab:Tabla1}
    \renewcommand{\arraystretch}{1.7}
    \resizebox{\textwidth}{!}{
        \begin{tabular}{|m{2.5cm}|m{2.3cm}|c|c|c|c|m{7cm}|}\toprule
            \centering\textbf{Dataset} & \centering\textbf{Source} & \centering\textbf{Observations} & \centering\textbf{Firms} & \centering\textbf{Frecuency} & \centering\textbf{Period} & \centering\textbf{Additional Information} \tabularnewline 
            \midrule
            \centering \textbf{Public Procurement Tenders}  & \centering \textbf{Spanish Goverment} &  381,000 &  100,000 & Daily & Aug 2019 - Jul 2024 & Data distinguishes between NGEU-funded and non-NGEU public contracts, including details such as sector (NACE 2-digit), number of awards, amounts, publication dates, etc.\\
            \hline
            
            \centering \textbf{New Lending Operations} & \centering \textbf{BBVA} & 5,090,000 & 318,000 & Daily & Aug 2019 - Jul 2024 & Includes firm-level corporate credit transactions: credit amount, weighted interest rate by credit amount, number of credits, long-term credit, short-term credit, and ICO credits.\\
            \hline
            
            \centering \textbf{Firm Characteristics} &  \centering \textbf{SABI} & - & 2.9 million & - & - & It includes company revenue (reported), number of employees, turnover, net capital, financial rating, and birth year (approximate age).\\
            \hline
            \rowcolor[rgb]{0.75,0.75,0.75}
            \centering \textbf{Final Aggregated Dataset} &  & 1,045,980 & 17,434 & Monthly & Aug 2019 - Jul 2024 & Balanced dataset. Final match combines credit performance, procurement awards, and socioeconomic features.\\ 
            \hline
            \rowcolor[rgb]{0.85,0.85,0.85}
            \centering \textbf{New Lending Operations} &  & 239,154 & 17,434 & - & - & \centering -\\ \tabularnewline
            \rowcolor[rgb]{0.85,0.85,0.85}
            \centering \textbf{Procurement awards} &  & 119,322 & 17,434 & - & - & \centering -\\ \tabularnewline
            \hline
            \rowcolor[rgb]{0.95,0.95,0.95}
            \centering \textbf{NGEU} &  & 21,350 & 2,062 & - & - & \centering - \\ 
            \tabularnewline
            \rowcolor[rgb]{0.95,0.95,0.95}
            \centering \textbf{Non-NGEU} &  & 1,024,630 & 17,282 & - & - & \centering - \\ 
            \tabularnewline
            \bottomrule
        \end{tabular}
    }
\end{table}


\section{Methodology}
\label{sec:Methodology}

\subsection{Public procurement bids and new credit operations}
\label{sec:MethodologyDuplicate1}


In estimating the effects of public procurement on new firm credit acquisition, it is important to acknowledge that a firm's response to being awarded a contract extends beyond the contemporaneous period. The dynamics of credit acquisition often persist after the initial award, as firms enter extensive project execution phases that require ongoing financing. This sustained demand for new credit is influenced by various factors, including the scale and duration of projects, which may necessitate securing additional funding at different stages of implementation. Furthermore, the need for new credit does not invariably materialize immediately following the contract award; in many instances, project execution may commence several months later. This temporal disconnect complicates the timing of credit behavior and underscores the necessity of estimating dynamic effects to capture the full scope of how public procurement contracts influence credit acquisition over time. 

We estimate the dynamic elasticity of new firm credit operations to public procurement bid awards by local projection panel regressions \citep{jorda2005estimation}. In particular, we estimate the cumulative growth rate of new firm credit before and after a firm is awarded a public procurement bid:

\begin{equation}
\label{fig:equation1}
\begin{split}
NL_{i,t+h} - NL_{i,t-1}=\alpha^{h}_{i} + \delta^{h}_{s,t} + \beta^{h} \cdot PROC_{i,t} + \lambda^{h} \cdot \textbf{X}_{i,t} + \theta^{h} \cdot \textbf{Y}_{i,t+h} + \epsilon^{h}_{i,t} \\
\forall_{ h \in \{-5, \ldots, 12\}}
\end{split}
\end{equation}

where the dependent variable is the difference between the logarithm of the cumulative of new lending operations (new credit) obtained by firm $i$ at month $t+h$ ($NL_{i,t+h}$) and the logarithm of new credit obtained at month $t-1$ ($NL_{i,t-1}$). The key regressor, $PROC_{i,t}$, is a dummy variable that takes value 1 when firm $i$ has been awarded at least one public procurement bid during month $t$, and 0 otherwise. Thus, coefficient $\beta^{h}$ directly represents the cumulative growth rate of new credit operations $h$ months before or after a firm is awarded a public procurement bid. The dynamic effects are investigated 12 months after the award\footnote{Due to sample reasons we limit the forward dynamic estimation one year after the award.}. Note that we allow time horizon $h$ to be both positive and negative. This is due to the very nature of the study; there are cases in which a single company emerges as the winner of a public procurement bid, and furthermore, this company has some prior certainty about its potential success in securing the contract. Under this scenario, we hypothesize about potential anticipatory effects in terms of credit acquisition by the company prior to the actual award, similar to \cite{gabriel2024credit}. Theoretically, the maximum period for executing such anticipation could span from the announcement of the bid to the granting of the contract. Considering the temporal difference between these two dates, the average duration in the sample used in this study is approximately four months. This is why we investigate these possible anticipatory effects up to a horizon of 4 months prior to the reference period ($t=-1$). In addition, we explore as a robustness check further anticipatory effects up to 10 months before the award.

We include firm ($\alpha^{h}_{i}$), and a combination of sector and time ($\delta^{h}_{s,t}$)\footnote{Sector-time fixed effects are a combination of time (month and year) and 2-digit NACE sector codes. Results are identical when applying first-letter NACE as sector indicator.} fixed effects. The inclusion of firm fixed effects is critical as it accounts for unique, time-invariant characteristics inherent to individual firms, such as management quality, market position, geographic location... The incorporation of sector-time fixed effects allows for the consideration of common trends and shocks that may influence firms within specific sectors over time. This dual fixed effects approach acknowledges that the impact of government demand on credit growth can be sector-dependent and may vary in response to broader economic conditions or policy changes. This includes potential heterogeneous monetary policy transmission across sectors, or energy and trade sector-specific shocks. 

Standard errors are clustered at the firm level to account for potential correlation of error terms within individual firms over time. This approach is essential because unobserved factors specific to each firm can influence multiple observations, leading to correlated errors that violate the independence assumption required for standard statistical inference. Clustering standard errors mitigates the risk of underestimating the true variability of estimates, which can result in misleadingly narrow confidence intervals and inflated t-statistics. This methodology enhances the robustness of the findings by providing a more accurate representation of the uncertainty associated with the estimated coefficients.

$\textbf{X}_{i,t}$ is a control vector that includes award-specific characteristics, containing amount of procurement, number of awards received each month, and project execution time\footnote{In case a firm is awarded more than one public procurement bid within a month, execution time of the award that represents the greatest import is selected. Furthermore, the inclusion and exclusion of execution time yields similar regression results.}, firm characteristics, including the log of age, number of employees and firm turnover, and the first lag of the main regressor (public procurement dummy) and the dependent variable ($NL_{i,t-1}$). Furthermore, we include an additional vector of controls $\textbf{Y}_{i,t+h}$ that is contemporaneous to the first element of the dependent variable. This vector uniquely includes the first lag of credit term (due to potential endogeneity issues), since we understand that the remaining term of credit is contemporaneously related to credit\footnote{This fact leaves the floor for a discussion on whether controls should be indexed at $t$ or $t-i$, instead of at $t+h$ in local projection regressions. In some cases, it is economically more pragmatic and realistic to include specific controls contemporaneous to the dependent variable in local projection settings.}.

\subsection{Public procurement bids and new credit operations: NGEU vs no-NGEU}
\label{sec:Methodology2}

Following the onset of the pandemic, a financing instrument was launched with the objective of supporting the European economic recovery anchored in a future that is digital, green, and resilient: the Next Generation EU (NGEU) instrument. This marked the introduction of a new type of public procurement financing, which began to coexist with existing mechanisms in 2021 and is set to conclude in 2026. Thus, the idea arises to analyze whether the effect on new business credit depends on the type of public procurement. Following the investigation of the effects of receiving public procurement contracts on new credit operations, our database facilitates an exploration of potential heterogeneous effects on credit contingent upon the type of public procurement contract. The differentiation between contracts financed by NGEU funds and those not financed by these funds (hereinafter referred to as no-NGEU) allows for an assessment of whether differential effects are present. To this end, we estimate the dynamic elasticity of new credit following the acquisition of both types of contracts by the following specification:

\begin{equation}
\label{fig:equation2}
\begin{split}
NL_{i,t+h} - NL_{i,t-1}=\alpha^{h}_{i} + \delta^{h}_{s,t} + \beta^{h} \cdot PROC^{{\scriptscriptstyle NGEU}}_{i,t} + \gamma^{h} \cdot PROC^{{\scriptscriptstyle NO-NGEU}}_{i,t} + \lambda^{h} \cdot \textbf{X}_{i,t} + \theta^{h} \cdot \textbf{Y}_{i,t+h} + \epsilon^{h}_{i,t} \\
\forall_{ h \in \{-5, \ldots, 12\}}
\end{split}
\end{equation}

where the main difference compared with specification (1) relies on the differentiation of two public procurement dummies. Particularly, $PROC^{{\scriptscriptstyle NGEU}}_{i,t}$ is a dummy variable that takes value 1 when firm $i$ has been awarded at least one NGEU-funded bid at month $t$, and 0 otherwise. Equivalently, $PROC^{{\scriptscriptstyle NO-NGEU}}_{i,t}$ takes value 1 when a firm has been awarded at least one bid not belonging to NGEU programs (rest of bids). Thus, coefficients $\beta^{h}$ and $\gamma^{h}$ will be interpreted as the dynamic elasticity of new firm credit after the award of NGEU and no-NGEU bids for every evaluated horizon $h$, respectively.

In addition, vector of controls $\textbf{X}_{i,t}$ includes a duplication of award-specific characteristics, one for each type of bid. In particular, we include and differentiate the amount of procurement, number of awards and project execution by the type of bid (NGEU vs no-NGEU). The set of fixed effects included, the estimation method and the horizon interval are identical to equation \ref{fig:equation1}.


\section{Results}
\label{sec:results1}

\label{sec:results_1}

In this section, we present the empirical findings on the relationship between public procurement contracts and firm-level new lending operations. We begin by analyzing the impact of all procurement bids on new lending, estimating the dynamic response of firm credit acquisition following contract awards. We then explore the heterogeneous effects across firm size, industry sectors, credit maturity, and supply chain position, assessing whether procurement-induced credit expansion varies based on firm and credit characteristics. Additionally, we examine the differential impact of NGEU versus no-NGEU procurement contracts, evaluating whether the Next Generation EU framework amplifies firm credit availability more effectively than traditional procurement. Finally, we discuss the transition from new lending to changes in the stock of credit, comparing our results to existing literature.

\subsection{The Effects of Public Procurement Bids on New Lending Operations}
\label{sec:results_1_duplicate}

In this section we show the results of regressions estimates presented in equation \ref{fig:equation1} on our full sample of public procurement firms (17.433 unique firms). 

\begin{figure}[h!]
    \setlength{\belowcaptionskip}{-5pt} 
    \centering
    \caption{\textbf{Response of New Credit Operations to Public Procurement Bids}} 
        \vspace{2pt} 
    \includegraphics[width=0.5\textwidth]{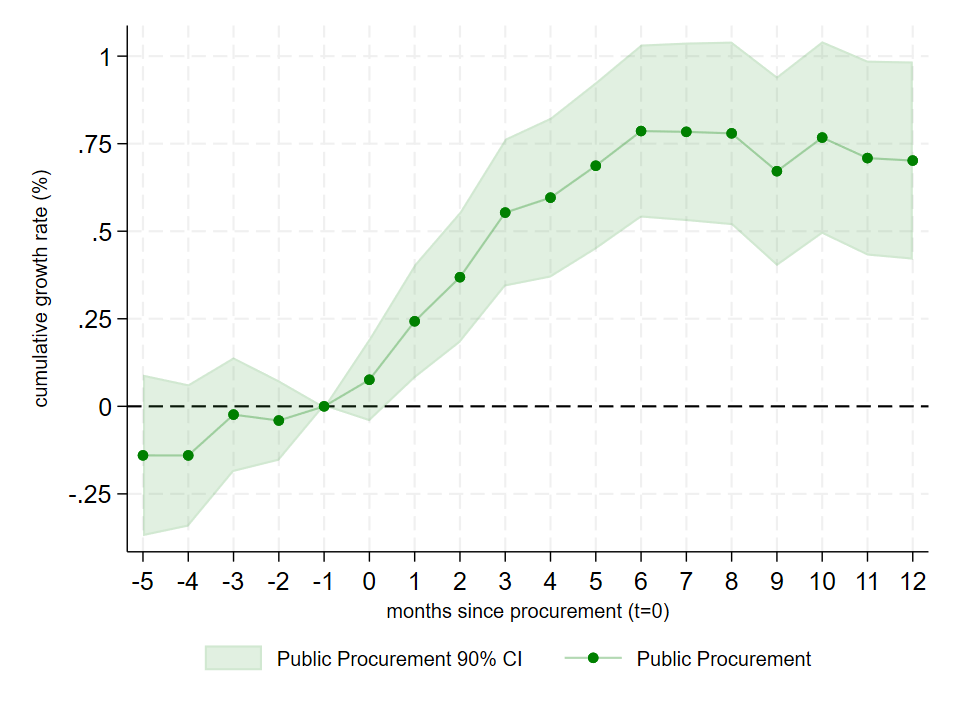}
    \label{fig:figura1}
    \begin{flushleft} 
        \scriptsize
        \vspace{-8pt} 

    \scriptsize
    \setlength{\parindent}{0pt}
    \setlength{\baselineskip}{8pt}
    Notes: The plot displays the estimated coefficient $\beta$ (green points) from regressions of equation \ref{fig:equation1} for each horizon $h$ relative to 1 month before public procurement awards, as well as its 90\% confidence bands (green shaded area). The estimated coefficient $\beta$ is interpreted as the cumulative growth rate of new credit operations $h$ months before or after procurement awards. The estimation includes firm and sector$\times$time fixed effects, and all standard errors are clustered at the firm level.\end{flushleft}
\end{figure}

Figure \ref{fig:figura1} illustrates the estimated cumulative impact of procurement awards on new lending, showing the coefficient estimate $\beta^{h}$ alongside its corresponding 90\% confidence intervals. The results suggest a 0.75\% cumulative increase in new credit operations within one year of winning a procurement contract. The expansion in new lending exhibits persistence , becoming statistically significant six months after the award and remaining robust across subsequent time horizons $h$.

The finding of persistence indicates that procurement contracts serve as an effective mechanism for enhancing firms' access to credit, not only at the time of the award but also in the long term. The sustained increase in credit demand suggests that firms leverage procurement contracts to expand borrowing capacity, likely due to the collateralization of future revenues from public sector contracts.

Our results show no evidence of anticipatory effect by firms (increasing and significant borrowing before awarded with the procurement contract). The new lending levels remain unchanged in the four months leading up to the contract award, confirming the absence of anticipatory effects. At the time of contract allocation ($h=0$), the estimated impact on credit is positive but not statistically significant. The credit response only becomes significant after six months and remains stable thereafter. This suggests that firms seek additional financing only after securing the contract and beginning project execution, rather than in anticipation of the award.

\clearpage

\vspace{0.5cm}
\vspace{-8pt} 

There is limited existing research estimating the dynamic elasticity of credit stock growth following public procurement awards. Notably, \cite{diGiovanni2022} find that winning a procurement contract in Spain between 2000 and 2013 is associated with a 5.5 percentage point increase in total firm credit stock. Similarly, \cite{gabriel2024credit}, using a back-of-the-envelope calculation, estimates that for Portuguese firms from 2009 to 2019, total credit stock grows by 3 percentage points one year after receiving a procurement contract.

These studies examine the impact of public procurement on changes in credit stock, which inherently incorporates amortization effects over time. As a result, while our findings on new lending operations offer a novel perspective, they are not directly comparable to the existing literature on credit stock dynamics. To bridge this gap, we apply a linear amortization assumption to each new credit issued, adjusting for its repayment schedule. This approach progressively reduces the value of new credit as it is repaid, enabling us to construct a new dependent variable that captures credit stock dynamics rather than new lending flows.

\begin{figure}[h!]
    \setlength{\belowcaptionskip}{-5pt} 
    \centering
    \caption{\textbf{Response of equivalent Credit Stock to Public Procurement Bids}} 
        \vspace{2pt} 
    \includegraphics[width=0.5\textwidth]{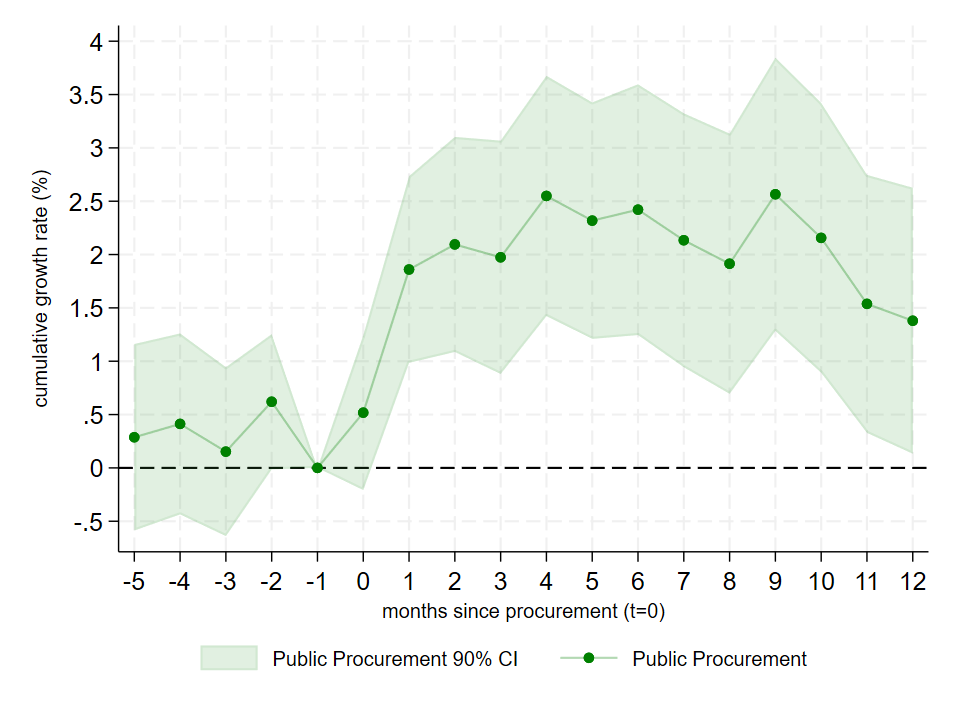}
    \label{fig:figura2}
    \begin{flushleft} 
        \footnotesize
        \vspace{-5pt} 
        Notes: the plot displays the estimated coefficient $\beta$ (green points) from regressions of equation \ref{fig:equation1} for each horizon $h$ relative to 1 month before public procurement awards, as well as its 90\% confidence bands (green shaded area). New credit operations stock is calculated by assuming linear credit repayments considering credit term. Thus, it represents an amortization-adjusted new credit operations. Estimated coefficient $\beta$ is interpreted as the cumulative growth rate of new credit operations stock $h$ months before or after procurement awards. The estimation includes firm and sector$\times$time fixed effects, and all standard errors are clustered at the firm level.
    \end{flushleft}
\end{figure}

Figure \ref{fig:figura2} presents the results from estimating equation \ref{fig:equation1} after applying the amortization adjustment. Our analysis indicates that one year after winning a procurement contract, the dynamic elasticity of credit stock is 1.5 percentage points. This estimate is lower than the 5.5 percentage point increase reported by \cite{diGiovanni2022} and the 3 percentage point increase reported by \cite{gabriel2024credit} in the case of Portugal.

Two important points deserve mention. First, the upper bound of our 90\% confidence interval reaches a cumulative growth of 3.0 percentage points, aligning closely with the findings of \cite{gabriel2024credit}. Second, our sample spans both the no-NGEU and NGEU periods, which may partly explain the discrepancy with the 5\% cumulative impact reported by \cite{diGiovanni2022} in Spain.

Within the first year, the elasticity of credit stock reaches a peak of approximately 3\% before gradually declining in the final three months. Furthermore, we find no evidence of anticipatory effects on credit stock, aligning with the findings of \cite{gabriel2024credit}. The differences observed between our results and those of previous studies may stem from variations in sample periods, firm characteristics, or differences in the composition of procurement contracts across datasets.

\clearpage

\vspace{0.5cm}

One could attribute those differences and the significant new credit impulse to the existence of the pandemic in our time sample period. To investigate that fact we conduct the same quantitative analysis but considering the concession of credits associated with COVID-related lines executed by the Credit Official Institute of Spain (ICO). These credit lines were designed to facilitate access to credit and improve the liquidity of businesses and self-employed individuals affected by the economic crisis caused by the pandemic \footnote{These lines of guarantees were established through two royal decree-laws: Royal Decree-Law 8/2020, approved on March 17, 2020, and Royal Decree-Law 25/2020, approved on July 3, 2020.}. By attending at credit product, we can identify which credits belong to such lines (COVID-related credits). In particular, we identify 5.616 credits categorized as such, which represent approximately 620 million euros. To this end, we estimate equation \ref{fig:equation1} while adjusting (subtracting) for new credit operations associated with these credits. As presented in Figure \ref{fig:figura11} in the Appendix, both the dynamic elasticity of new credit operations (panel a) and new credit stock (panel b) are similar to the case of no correction.

\subsubsection{Firm characteristics: activity sector and size}
\label{sec:results}



This section examines how the impact of public procurement on firm credit varies by activity sector and firm size. The results show that firms in the manufacturing sector experience the largest credit expansion, while firms in wholesale trade and retail exhibit a strong but short-lived response. In contrast, administrative and professional services firms show no significant effect, and construction firms display a moderate but sustained increase in credit availability. Regarding firm size, smaller firms benefit more from procurement awards, with credit elasticity reaching 1.25\% one year after the award, compared to 0.5\% for larger firms.

\textbf{Heterogeneity by Industry Sector}

Public procurement programs often prioritize certain industries based on economic and policy objectives, influencing the credit response across sectors. In our sample, 70\% of firms operate in wholesale trade and retail (25.3\%), construction (16.2\%), administrative and professional activities (15.3\%), and manufacturing (12\%). 

\begin{figure}[hbt!]
    \caption{\textbf{Response of New Credit Operations to Public Procurement Bids by Firm Activity Sector}} 
    \vspace{2pt} 
    \label{fig:figura3}
\centering
\begin{subfigure}{.3\linewidth}
  \includegraphics[width=\linewidth]{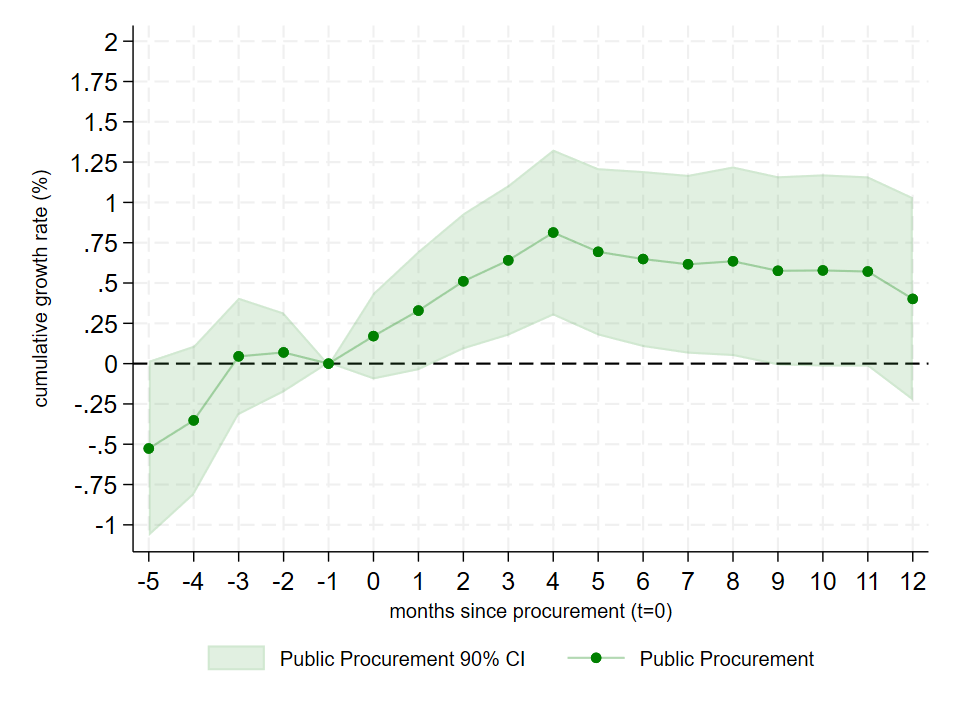}
  \caption{\textbf{Wholesale trade and retail}} 
  \label{subfig:MLEDdet}
\end{subfigure}\hspace{-2pt} 
~ 
\begin{subfigure}{.3\linewidth}
  \includegraphics[width=\linewidth]{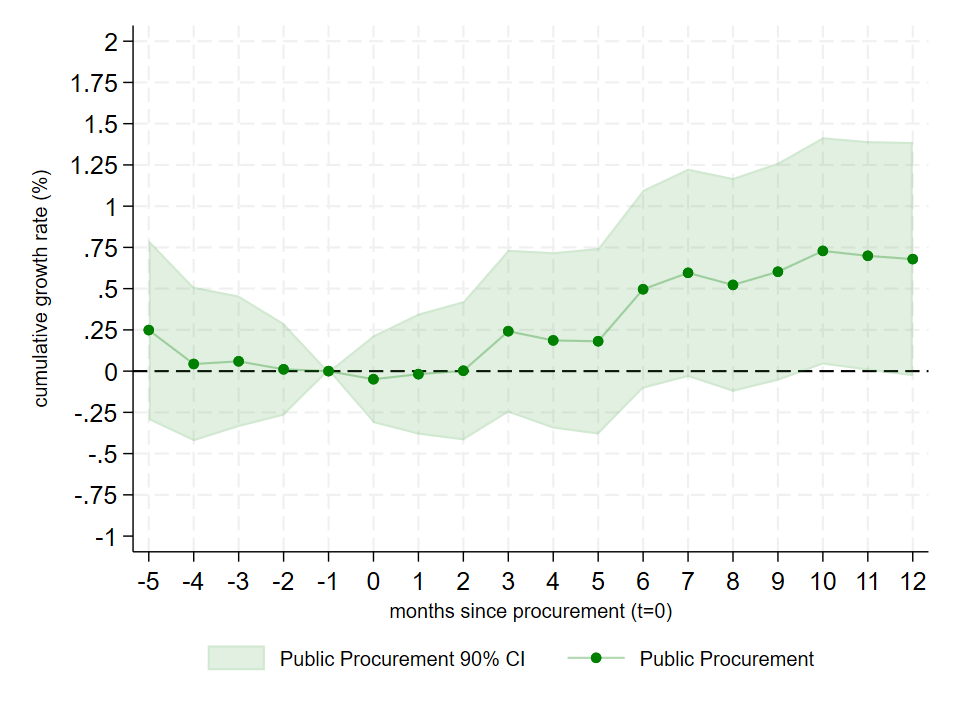}
  \caption{\textbf{Construction}} 
  \label{energydetPSK}
\end{subfigure}

\medskip 

\begin{subfigure}{.3\linewidth}
  \includegraphics[width=\linewidth]{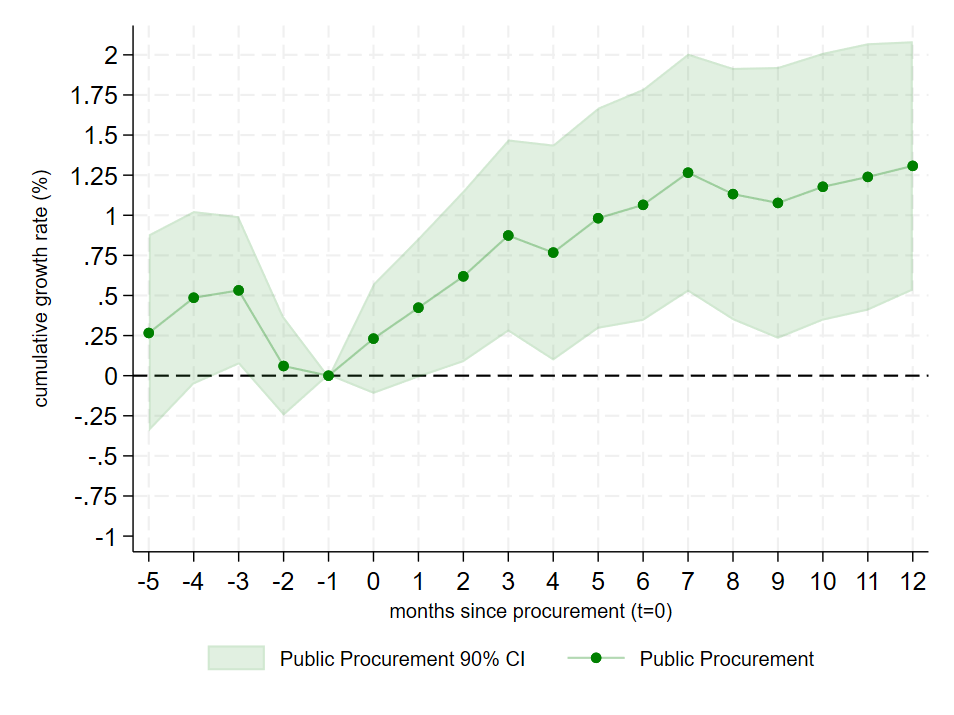}
  \caption{\textbf{Manufacturing industry}} 
  \label{velcomp}
\end{subfigure}\hspace{3pt} 
\begin{subfigure}{.3\linewidth}
  \includegraphics[width=\linewidth]{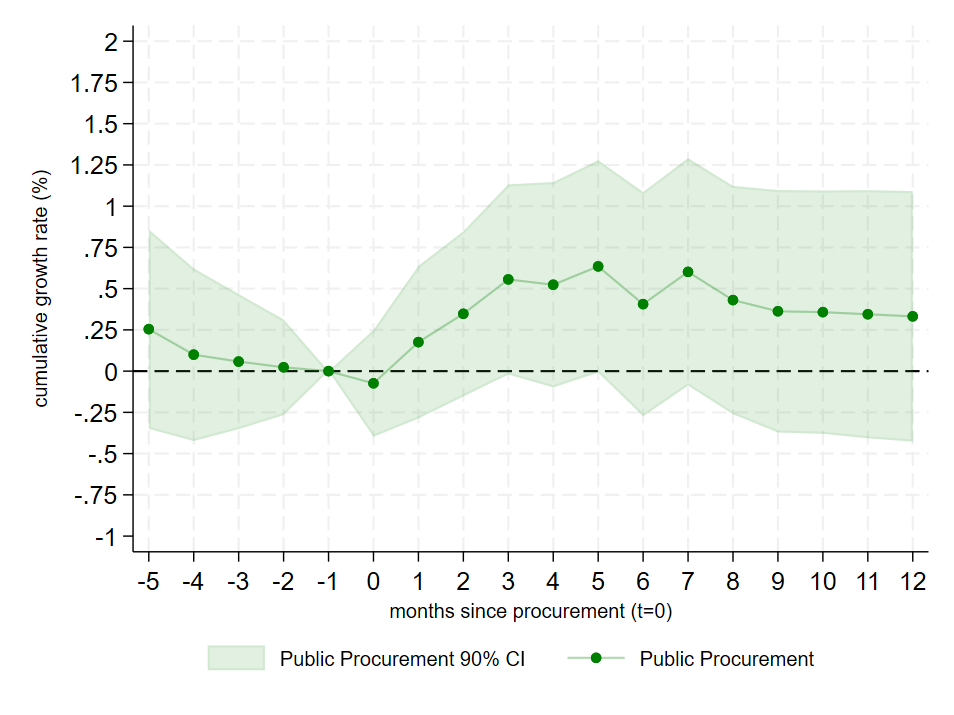}
  \caption{\textbf{Administrative and professional activities}} 
  \label{estcomp}
\end{subfigure}

    \vspace{-8pt} 
    \begin{flushleft} 
    \scriptsize
    \setlength{\parindent}{0pt}
    \setlength{\baselineskip}{8pt}
    
        \scriptsize{
        Notes: The plots display the estimated coefficient $\beta$ (green points) from regressions of equation \ref{fig:equation1} for each horizon $h$ relative to 1 month before public procurement awards, as well as its 90\% confidence bands (green shaded area). Panel (a) shows the results for a sub-sample of firms dedicated to wholesale trade and retail activities (first letter NACE code G), and panel (b) for construction firms (F), panel (c) for manufacturing firms (C) and (d) for the set of administrative and support service activities, and professional, scientific and technical activities (letters N \& M, respectively). These sectors are the most represented in the sample of companies used in this paper. The estimated coefficient $\beta$ is interpreted as the cumulative growth rate of new credit operations (by activity sector) $h$ months before or after procurement awards. The estimation includes firm and sector$\times$time fixed effects, and all standard errors are clustered at the firm level.}
    \end{flushleft}
    \vspace{-8pt}
\end{figure}

In this section, we examine whether the dynamic elasticity of new credit depends on the company's sector of activity by estimating the regressions corresponding to equation \ref{fig:equation1}, distinguishing between the most representative sectors in the sample. Figure \ref{fig:figura3} illustrates that companies operating in the manufacturing industry experience the largest credit boost (an accumulated increase exceeding 1.25\% in the first year). Similarly, construction companies receive the second-largest impulse in the first year (0.75\%). In contrast, wholesale and retail, as well as administrative and professional activities, experience a more modest impulse, which is not significant at the 90\% confidence level one year after the award. Thus, government-dependent sectors (construction and manufacturing) experience the largest credit boosts, while firms in less government-dependent industries exhibit more muted responses.

\clearpage

\textbf{Heterogeneity by Firm Size}

The impact of public procurement also depends on firm size, as smaller firms often face greater financial constraints and benefit more from government contracts. While awarding contracts to small enterprises may involve higher short-term costs for both the public and private sectors, the long-term benefits include increased investment, business expansion, and job creation. Thus, the fact that small enterprises experience a greater credit impulse may indicate increased investment on their part, leading to subsequent enhanced long-term growth, provided that the size of public contracts is not reduced and that the ease of procedures is improved \citep{aguilar2023primera,diGiovanni2022}.

\begin{figure}[h!]
    \centering
    \caption{\textbf{Response of New Credit Operations to Public Procurement Bids by Firm Size}} 
    \vspace{2pt} 
    \label{fig:figura4}
    \begin{subfigure}[b]{0.45\textwidth} 
        \centering
        \includegraphics[width=\textwidth]{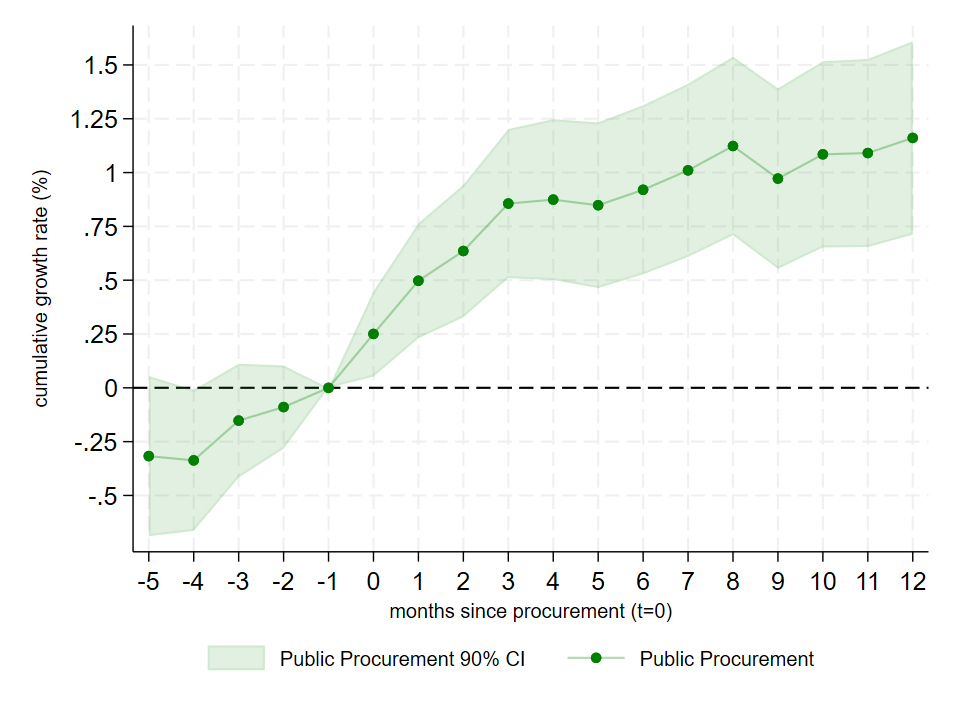} 
        \caption{\textbf{Small firms}} 
        \label{fig:graph1}
    \end{subfigure}
    \hspace{-5pt} 
    \begin{subfigure}[b]{0.45\textwidth} 
        \centering
        \includegraphics[width=\textwidth]{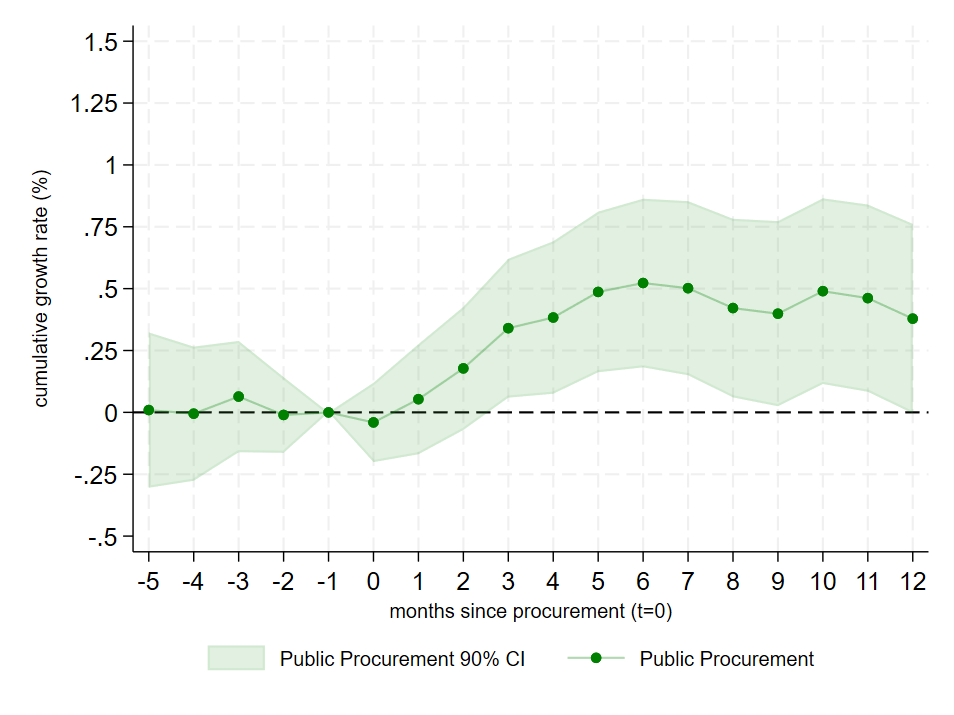} 
        \caption{\textbf{Large firms}} 
        \label{fig:graph2}
    \end{subfigure}

     \vspace{-8pt} 
    \begin{flushleft} 
    \scriptsize
    \setlength{\parindent}{0pt}
    \setlength{\baselineskip}{8pt} 
        \scriptsize{
        Notes: The plots display the estimated coefficient $\beta$ (green points) from regressions of equation \ref{fig:equation1} for each horizon $h$ relative to 1 month before public procurement awards, as well as its 90\% confidence bands (green shaded area). The following sub categorization of firms relies on firm turnover measured in euros. In particular, the median of sample turnover is the division threshold (2 million euros). Panel (a) shows the results for a sub-sample of firms categorized as small (turnover lower or equal than the median), and (b) for a sub-sample of large firms (turnover greater than sample median). Estimated coefficient $\beta$ is interpreted as the cumulative growth rate of new credit operations $h$ months before or after procurement awards. The estimation includes firm and sector$\times$time fixed effects, and all standard errors are clustered at the firm level}
    \end{flushleft}
\end{figure}

To this end, we categorize companies based on their turnover level\footnote{Note that the last turnover figure available is selected for each company. In cases when turnover is missing in subsequent years, it is remained constant. The analysis yields similar results when the sample median of employees number is utilized as the division threshold.}, dividing the sample according to the sample median (€2 million). Figure \ref{fig:figura4} presents the results of estimating the regressions of equation \ref{fig:equation1} for both groups of companies. The results indicate that the dynamic response of credit following the award of public procurement contracts depends on firm size; small companies experience an increase in new credit of approximately 1.25\% in cumulative terms one year later, whereas larger companies see their credit boosted significantly but by approximately 0.5\%. In neither case are significant anticipatory effects observed. This heterogeneity highlights the role of firm characteristics in shaping the effectiveness of procurement in firm financing and the broader macroeconomic impact of NGEU funds in supporting economic recovery.

\clearpage

\vspace{0.5cm}

\subsubsection{The Firm position on the Production Value Chain: Proximity to the Consumer}
\label{sec:results_2}

Recent research has stressed the potential usefulness of considering production networks as an important determinant of the transmission of policy shocks. There are some examples on this effects by monetary policy (see for example, (\cite{ozdagli2017,ghassibe2021}) for early empirical findings along these lines). More recently \cite{buda2023} shown that in the case of Spain, monetary policy shocks have had a strong and quick response in those downstream activities closer to the final consumers in response to an homogeneous negative shock.

In the context of fiscal policy, public procurement affects firms at all stages of the supply chain, allowing governments to strategically allocate contracts to support firms operating either closer to or further from the final consumer. Our analysis focuses on evaluating how the impact of procurement on new credit varies based on a firm's position within the supply chain.

To measure this, we rely on the upstreamness metric developed by \cite{antras2012} and adapted by \cite{buda2023}. This metric provides a quantitative assessment of an industry's position within the global production network, capturing the relative distance of a firm’s industry from final consumption. A higher upstreamness score indicates that a firm operates in the earlier stages of production, supplying inputs to other firms rather than selling directly to consumers. Conversely, a lower upstreamness score suggests that a firm is positioned closer to the end of the supply chain, directly engaging in final goods and services. This approach allows us to examine whether procurement contracts disproportionately benefit firms at specific points in the production process and how these effects translate into changes in credit availability.

By establishing a preliminary correspondence between industries in the Input-Output tables and the list of 2-digit CNAE sectors, we matched the 87 CNAE sectors with each upstreamness indicator, utilizing evidence from both studies. Specifically, this indicator ranges in value from 1 to 4, where 1 indicates that the firm operates as close as possible to the consumer and 4 indicates operation from the most distant point. Based on this indicator, we create first  a binary variable that takes value 0 for firms classified as downstream, or relatively close to the consumer, if the metric is equal to or less than 2.2. Conversely, the variable takes value 1 and firms are classified as upstream, or distant from the final consumer, if the metric exceeds this threshold, following the approach of \cite{antras2012} and \cite{buda2023}. \footnote{This method of categorizing companies classifies those operating in the construction sector, textile and electronic manufacturing, retail and wholesale, education, and social services as downstream, while the upstream group includes companies engaged in the supply of electricity and water, professional and administrative activities, chemical manufacturing, and mining, among many others. This classification of companies represents, in aggregate terms, a division of the total amount awarded in public procurement bids amounting to €71 trillion for downstream companies and €21 trillion for upstream companies. Consequently, we generated an initial group of 11,353 unique firms that are relatively close to the consumer and another group of 5,650 firms that are further removed from the final consumption phase. Table \ref{tab:Tabla3} in the Appendix shows the upstreamness metric and binary indicator for all CNAE 2-digit codes and descriptions.}
\vspace{0.5cm}

\begin{figure}[hbt!]
    \caption{\textbf{Response of New Credit Operations to Public Procurement Bids by Upstreamness}} 
    \vspace{2pt} 
    \label{fig:figura5}
\centering
\begin{subfigure}{.3\linewidth}
  \includegraphics[width=\linewidth]{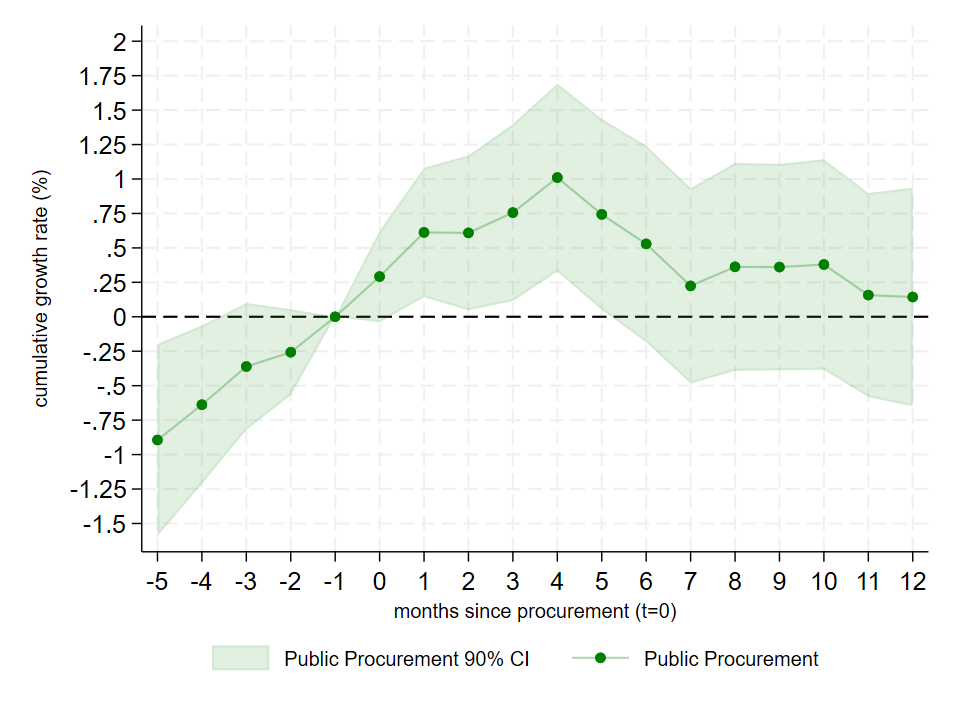}
  \caption{\textbf{Very downstream}} 
  \label{MLEDdet}
\end{subfigure}\hspace{-2pt} 
~ 
\begin{subfigure}{.3\linewidth}
  \includegraphics[width=\linewidth]{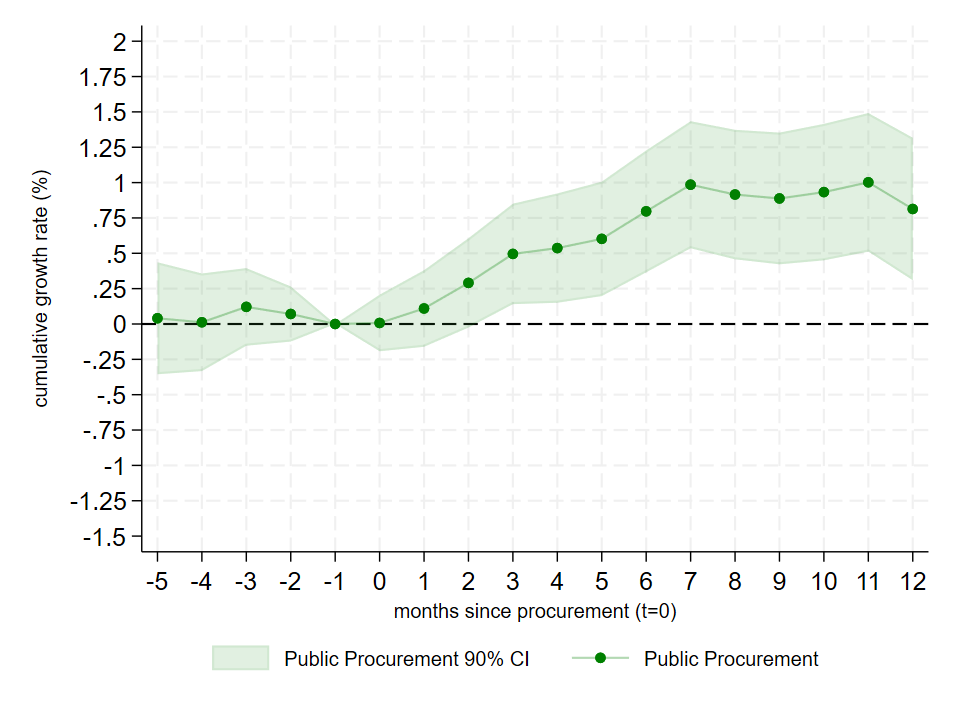}
  \caption{\textbf{Less downstream}} 
  \label{energydetPSK2}
\end{subfigure}

\medskip 

\begin{subfigure}{.3\linewidth}
  \includegraphics[width=\linewidth]{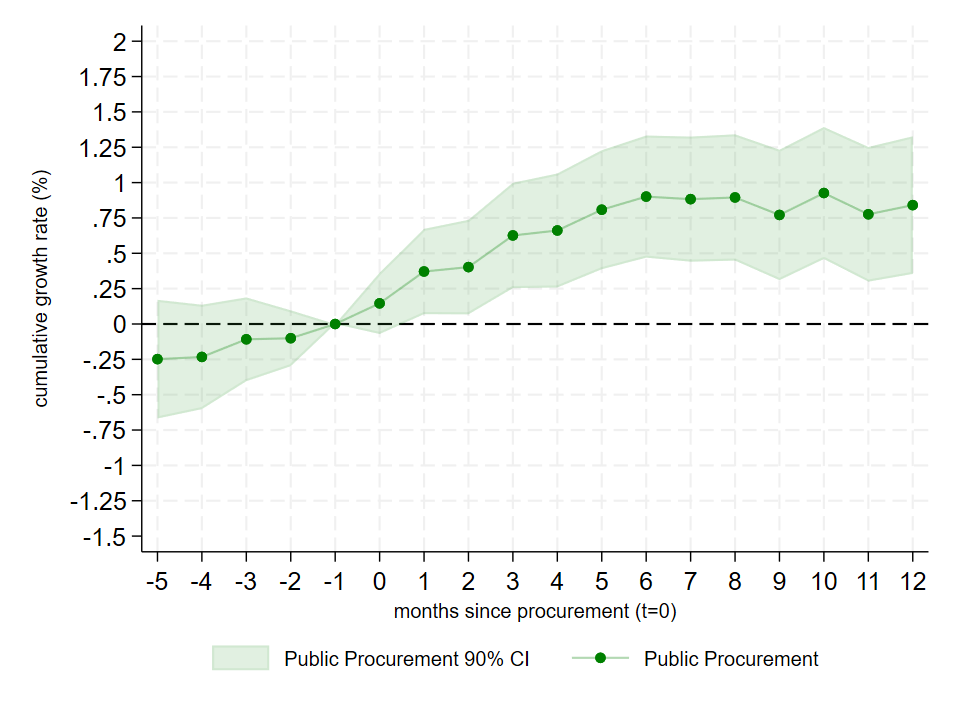}
  \caption{\textbf{Less upstream}} 
  \label{velcomp2}
\end{subfigure}\hspace{3pt} 
\begin{subfigure}{.3\linewidth}
  \includegraphics[width=\linewidth]{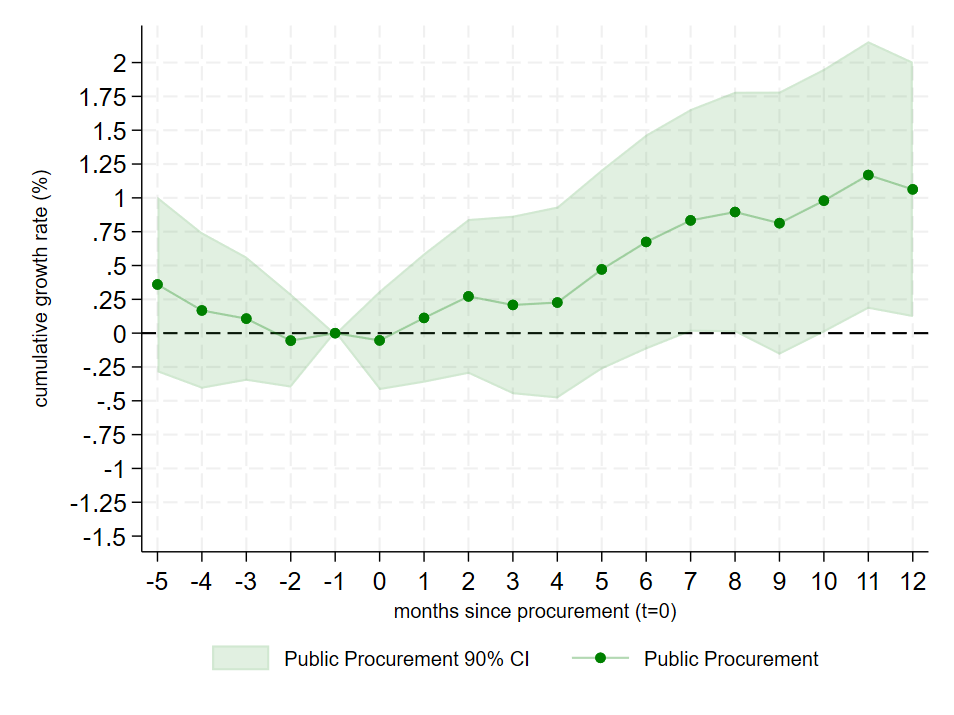}
  \caption{\textbf{Very upstream}} 
  \label{estcomp2}
\end{subfigure}

    \vspace{-8pt} 
    \begin{flushleft} 
    \scriptsize
    \setlength{\parindent}{0pt}
    \setlength{\baselineskip}{8pt}
    
        \scriptsize{
        Notes: The plots display the estimated coefficient $\beta$ (green points) from regressions of equation \ref{fig:equation1} for each horizon $h$ relative to 1 month before public procurement awards, as well as its 90\% confidence bands (green shaded area). Panel (a) shows the results for the case of restricting the sample to firms categorized as proximate to the consumer (downstream), and panel (b) for the case of firms far from consumer (upstream). The categorization is based on \cite{antras2012}, and further developed for the case of Spain by \cite{buda2023}. Estimated coefficient $\beta$ is interpreted as the cumulative growth rate of new credit operations $h$ months before or after procurement awards. The estimation includes firm and sector$\times$time fixed effects, and all standard errors are clustered at the firm level.
        }
    \end{flushleft}
    \vspace{-8pt}
\end{figure}

After classifying individual firms based on their position within the production network, we estimate equation \ref{fig:equation1} separately for upstream and downstream firms (Figure \ref{fig:figura12} in the Appendix). Both groups of firms experience on average a significant and positive credit impulse of 0.75\% and 1\%, for downstream and upstream firms, respectively. However, we do not observe highly latent differences, which could be influenced by the fact that the bulk of firms is positioned in the middle of the upstreamness score.

This classification of firms based on upstreamness relies on a threshold of \(2.2\), which is somewhat arbitrary and may misclassify certain industries. For instance, civil engineering (CNAE 42) is categorized as downstream, similar to retail trade, even though it could reasonably be considered upstream. To address this limitation, we introduce an alternative classification (Figure \ref{fig:figura5}) that divides firms into four groups based on their position in the supply chain: very downstream (\(\leq 1.77\)), less downstream (\(1.82 \leq 2.19\)), less upstream (\(2.2 \leq 2.78\), including civil engineering), and very upstream (\(\geq 2.8\)).

Figure \ref{fig:figura5} indicates how the results show clear differences in credit response across these groups. Very downstream firms exhibit an immediate credit increase following procurement awards, while very upstream firms experience a delayed but sustained response. Firms in the moderate upstreamness categories behave similarly, with significant and lasting credit expansion over time. This refined classification provides a more nuanced understanding of how procurement awards impact credit availability at different stages of the supply chain.

These results partially match with activity sector-specific responses. Retail and wholesale firms new credit react quicker but fades out before the end of the first complete year, similar to very downstream firms. Also, firms operating in the manufacturing industry display persistent effects, similar to less downstream and upstream firms.

\begin{figure}[h!]
    \setlength{\belowcaptionskip}{-5pt} 
    \centering
    \caption{\textbf{Response of New Credit Operations to Public Procurement Bids: Very Downstream vs Very Upstream}} 
        \vspace{2pt} 
    \includegraphics[width=0.5\textwidth]{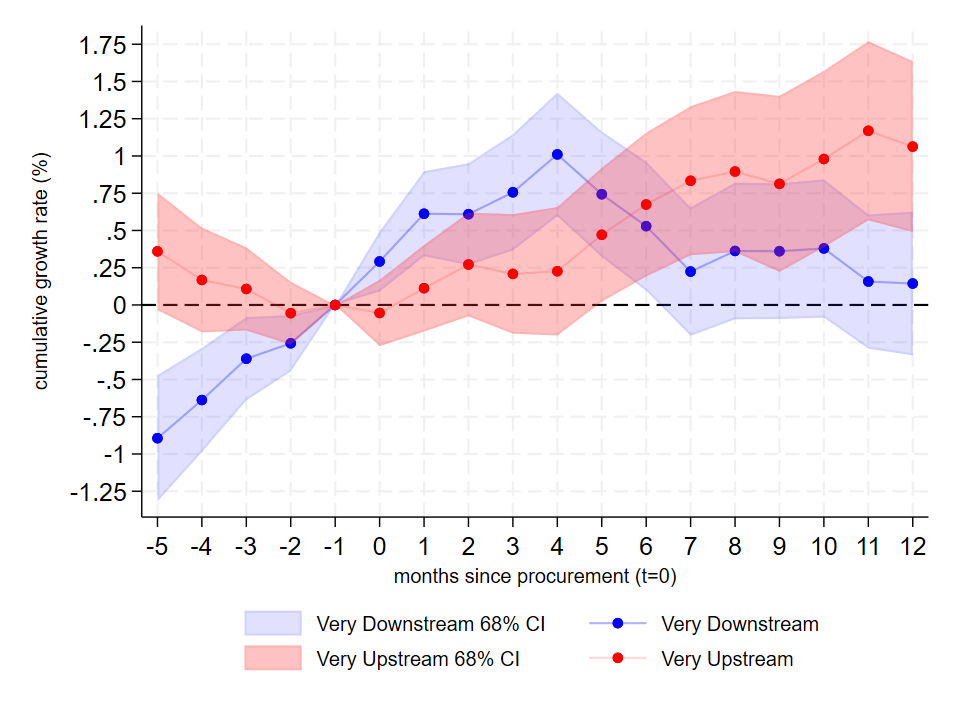}
    \label{fig:figura6}
   
    \begin{flushleft} 
    \scriptsize
    \setlength{\parindent}{0pt}
    \setlength{\baselineskip}{8pt} 
    
        \vspace{-5pt} 
        \scriptsize{
       Notes: The plot displays the estimated coefficient $\beta$ from regressions of equation \ref{fig:equation1} for each horizon $h$ relative to 1 month before public procurement awards, as well as its 68\% confidence bands (shaded areas), for the case of very downstream firms (blue points) and very upstream (red points). Blue estimates shows the results for the case of restricting the sample to firms categorized as very proximate to the consumer (upstreamness indicator below or equal 1.77, excepting wholesale trade 2-digit CNAE sectors), and red estimates for the set of very upstream firms (greater or equal than 2.8). The categorization is based on \cite{antras2012}, and further developed for the case of Spain by \cite{buda2023}. The estimated coefficient $\beta$ is interpreted as the cumulative growth rate of new credit operations (by extreme upstreamness degree) $h$ months before or after procurement awards. The estimation includes firm and sector$\times$time fixed effects, and all standard errors are clustered at the firm level.}
    \end{flushleft}
\end{figure}

Moreover, Figure \ref{fig:figura6} exhibits a more in detailed analysis on firms on the the extremes of the production network; very downstream vs very upstream firms. At the 68\% confidence level, the response in the initial months by companies closer to the consumer is significantly greater than that of those farther away, whereas one year after the procurement, the trends reverse, and it is the upstream companies that experience a higher and more persistent boost.

\clearpage

\subsubsection{Short vs long-term credit}
\label{sec:results_2_alt}

The nature and characteristics of the public tender may also influence the volume and amount of new credit obtained following the award. The acquisition of a public contract may require continuous access to new credit over different time frames. Consequently, the effects of public procurement on credit dynamics may vary depending on the maturity of the contracts.

Our dataset enables us to differentiate new credit obtained by firms at different maturity. To do this, the dependent variable in equation \ref{fig:equation1} is differentiated into new short-term credit, defined as credit with a maturity of one year or less, and new long-term credit, with a maturity exceeding one year. This distinction allows us to analyze whether the carry-over effect of public procurement awards on credit varies across different maturity periods.

The classification of credit in this way is based on its term, measured in days within our dataset. Specifically, the median and mean credit terms are 85 and 273 days, respectively, while the first and third percentiles of the distribution correspond to 38 and 1,826 days, respectively. In aggregate terms, short-term credit constitutes 68\% of total new credit, while long-term credit accounts for the remaining 32\%.

\begin{figure}[h!]
    \centering
    \caption{\textbf{Response of New Credit Operations to Public Procurement Bids by Credit Maturity}} 
    \vspace{2pt} 
    \label{fig:figura7}
    \begin{subfigure}[b]{0.45\textwidth} 
        \centering
        \includegraphics[width=\textwidth]{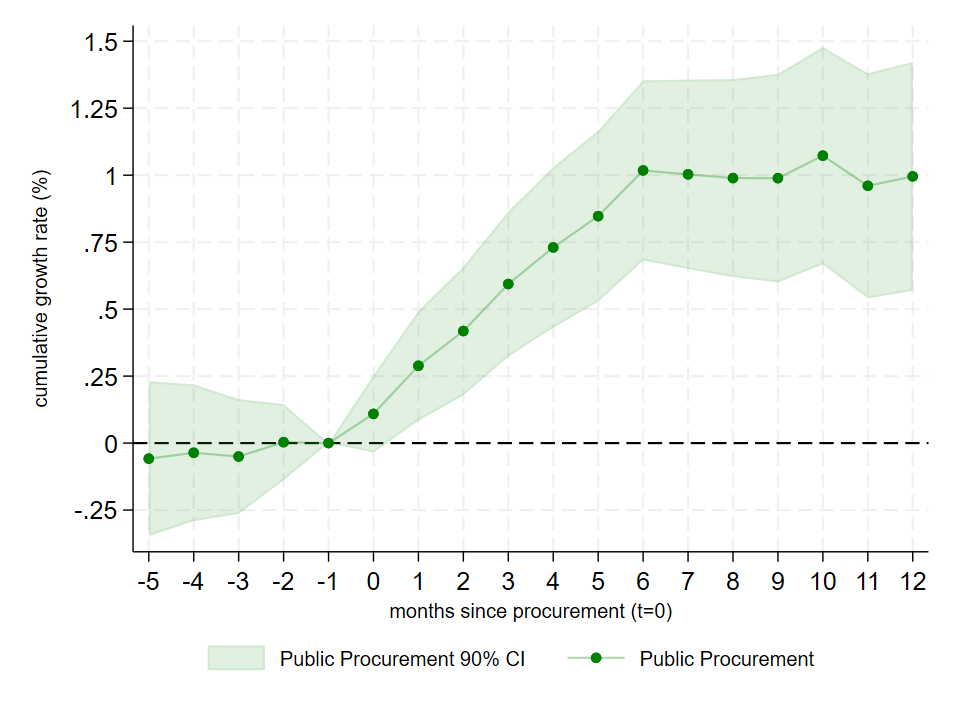} 
        \caption{\textbf{Short-term credit}} 
        \label{fig:graph1a}
    \end{subfigure}
    \hspace{-5pt} 
    \begin{subfigure}[b]{0.45\textwidth} 
        \centering
        \includegraphics[width=\textwidth]{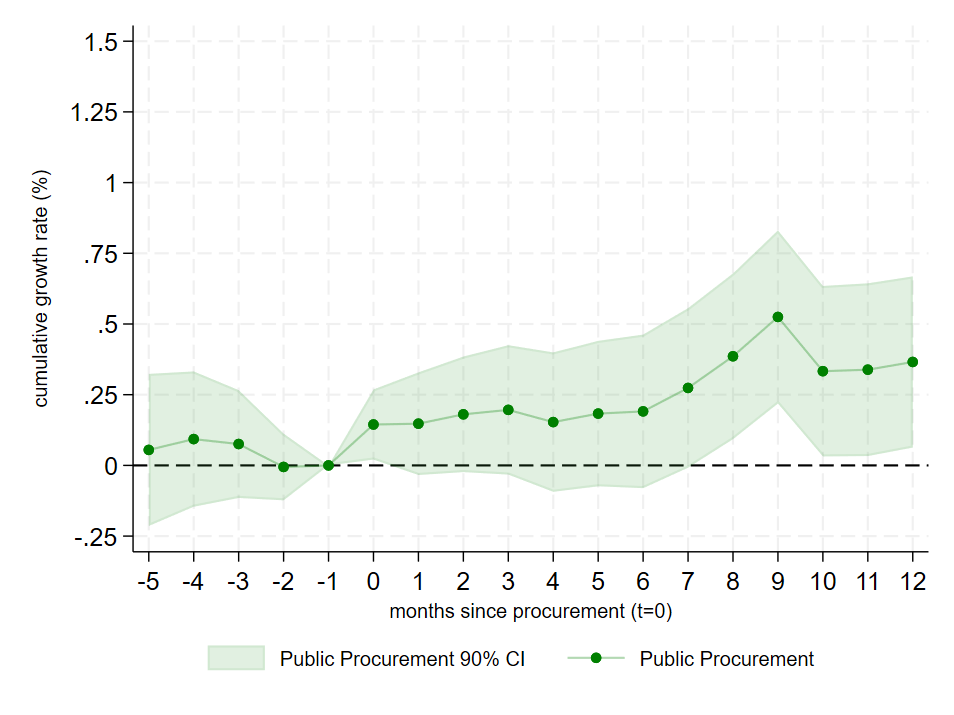} 
        \caption{\textbf{Long-term credit}} 
        \label{fig:graph2a}
    \end{subfigure}
    
    \begin{flushleft} 
    \scriptsize
    \setlength{\parindent}{0pt}
    \setlength{\baselineskip}{8pt} 
    
        \vspace{-5pt} 
        \scriptsize{
        Notes: The plots display the estimated coefficient $\beta$ (green points) from regressions of equation \ref{fig:equation1} for each horizon $h$ relative to 1 month before public procurement awards, as well as its 90\% confidence bands (green shaded area). Panel (a) shows the results for the case of the dependent variable being firm short-term new credit (maturity below or equal to 1 year), and panel (b) for the case of long-term new credit (maturity above 1 year). Estimated coefficient $\beta$ is interpreted as the cumulative growth rate of new credit operations (short or long term) $h$ months before or after procurement awards. The estimation includes firm and sector$\times$time fixed effects, and all standard errors are clustered at the firm level}
    \end{flushleft}
\end{figure}

Figure \ref{fig:figura7} presents the estimated coefficients $\beta^{h}$ for both dependent variables. Panel (a) displays the effects on new short-term credit, whereas panel (b) illustrates the effects on new long-term credit. The estimates indicate that public procurement awards have a more pronounced influence on short-term credit, with an increase of approximately 1\% observed one year after the contract is awarded, in contrast to a relatively modest increase of 0.3\% for long-term credit. This substantial rise in short-term credit may be attributed to the ongoing need for financing to support and sustain the execution of public procurement projects, which frequently necessitate reliance on shorter-maturity credit.

\clearpage

\vspace{0.5cm}



\subsection{Public procurement bids and new credit operations: NGEU vs no-NGEU}

The sample period analyzed in this study coincides with the emergence of the Next Generation EU funds. Our database further distinguishes whether any public tender has been financed with these funds or, on the contrary, has been financed by other European structural funds or national funds. Thus, we estimate the set of regressions exposed in equation \ref{fig:equation2}, differentiating between NGEU and no-NGEU awards. 

\begin{figure}[h!]
    \setlength{\belowcaptionskip}{-5pt} 
    \centering
    \caption{\textbf{Response of New Credit Operations to Public Procurement Bids: NGEU vs no-NGEU}} 
        \vspace{2pt} 
    \includegraphics[width=0.5\textwidth]{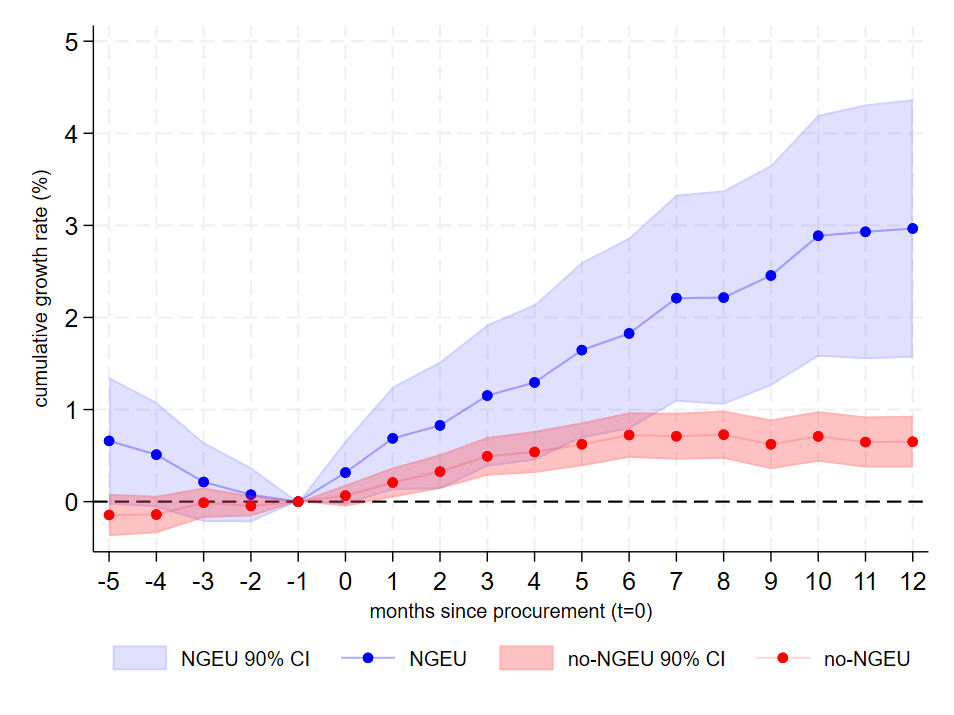}
    \label{fig:figura8}
 
    \begin{flushleft} 
    \scriptsize
    \setlength{\parindent}{0pt}
    \setlength{\baselineskip}{8pt} 
    
        \vspace{-5pt} 
        \scriptsize{
        Notes: the plot displays the estimated coefficients $\beta$ (blue points) and $\gamma$ (red points) from regressions of equation \ref{fig:equation2} for each horizon $h$ relative to 1 month before NGEU and no-NGEU public procurement awards, respectively, as well as their 90\% confidence bands (blue and red shaded areas for NGEU and no-NGEU, respectively). Estimated coefficients $\beta$ and $\gamma$ are interpreted as the cumulative growth rate of new credit operations $h$ months before or after NGEU and no-NGEU procurement awards, respectively. Note that no-NGEU public procurement bids refer to those different from NGEU-funded bids. The estimation includes firm and sector$\times$time fixed effects, and all standard errors are clustered at the firm level.}
    \end{flushleft}
\end{figure}

Figure \ref{fig:figura8} presents the results of estimating the parameters $\beta^{h}$ and $\gamma^{h}$, which represent the dynamic elasticity of new corporate credit before and after the award of tenders financed by NGEU and no-NGEU funds, respectively. A clear heterogeneity in the dynamic impact is observed; new credit increases by 3\% one year after NGEU awards in cumulative terms, while no-NGEU funds are associated with a 0.75\% growth in new credit, a result similar to the effect of all public procurement bids\footnote{The number of no-NGEU awards in the sample is considerably larger than NGEU awards}. Note that in neither of the two cases are significant anticipatory effects observed in terms of access to new credit.

\clearpage

Moreover, the availability of credit data allows us to examine the potential differential effects of NGEU-funded versus no-NGEU-funded public procurement bids on credit maturity. To analyze these effects, we estimate the regressions presented in equation \ref{fig:equation2}, this time distinguishing between short-term and long-term credit as separate dependent variables. 

\begin{figure}[h!]
    \centering
    \caption{\textbf{Response of New Credit Operations to Public Procurement Bids by Credit Maturity: NGEU vs no-NGEU}} 
    \vspace{2pt} 
    \label{fig:figura9}
    \begin{subfigure}[b]{0.45\textwidth} 
        \centering
        \includegraphics[width=\textwidth]{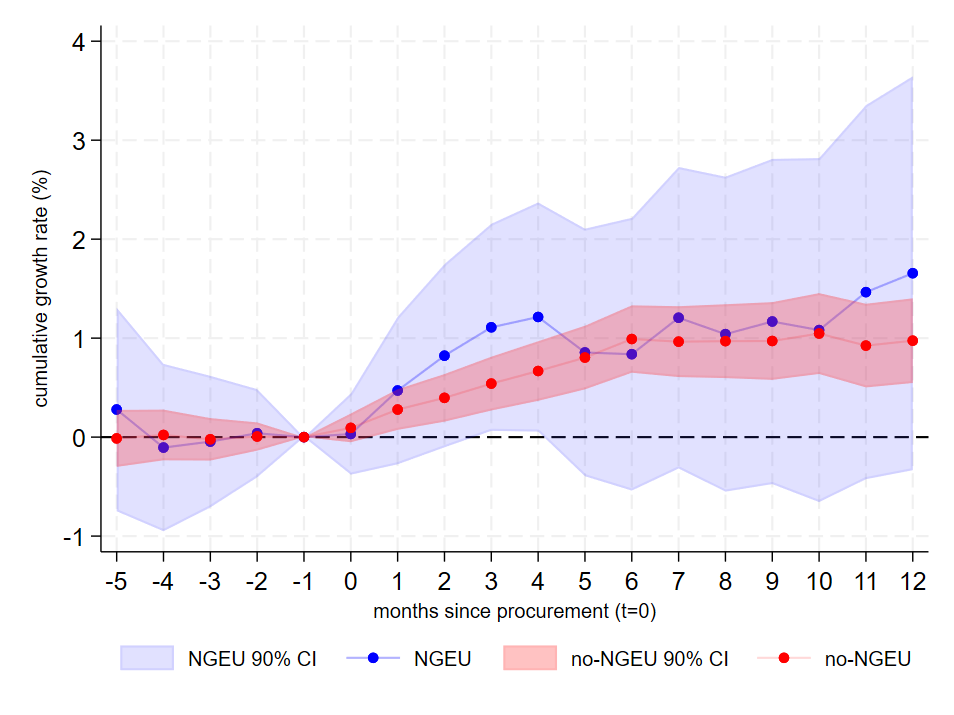} 
        \caption{\textbf{Short-term credit}} 
        \label{fig:graph1a_1}
    \end{subfigure}
    \hspace{-5pt} 
    \begin{subfigure}[b]{0.45\textwidth} 
        \centering
        \includegraphics[width=\textwidth]{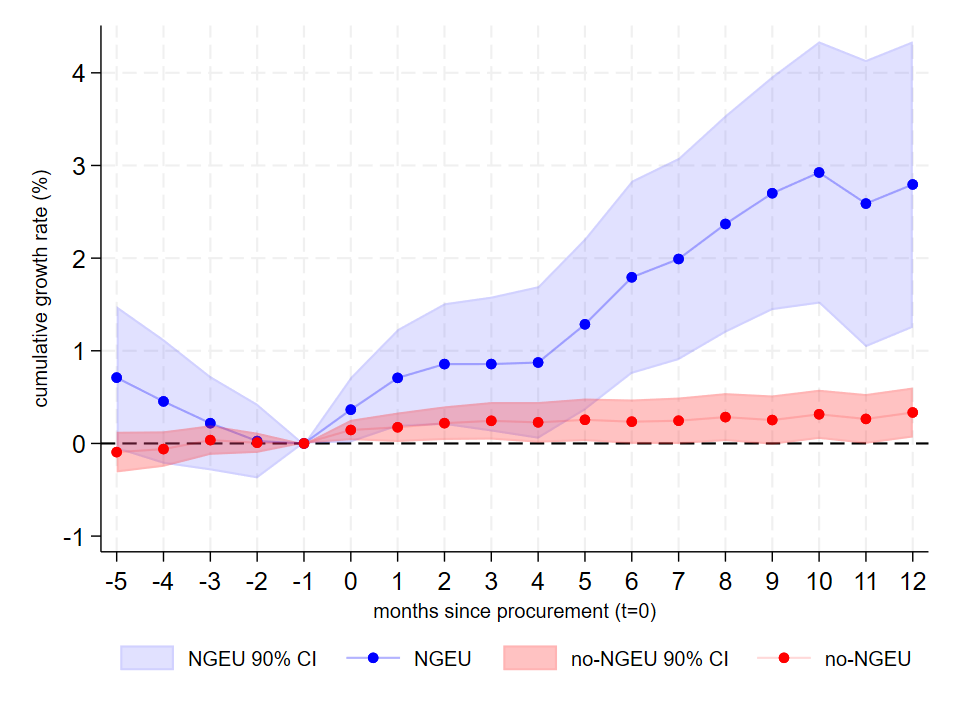} 
        \caption{\textbf{Long-term credit}} 
        \label{fig:graph2_1}
    \end{subfigure}
    
    \vspace{-5pt} 

    \begin{flushleft} 
    \scriptsize
    \setlength{\parindent}{0pt}
    \setlength{\baselineskip}{8pt} 
    
        \vspace{-5pt} 
        \scriptsize{
        Notes: the plots display the estimated coefficients $\beta$ (blue points) and $\gamma$ (red points) from regressions of equation \ref{fig:equation2} for each horizon $h$ relative to 1 month before NGEU and no-NGEU public procurement awards, respectively, as well as their 90\% confidence bands (blue and red shaded areas for NGEU and no-NGEU, respectively). Panel (a) shows the results for the case of the dependent variable being firm short-term new credit (maturity below or equal to 1 year), and panel (b) for the case of long-term new credit (maturity above 1 year). Estimated coefficients $\beta$ and $\gamma$ are interpreted as the cumulative growth rate of new credit operations (short or long term) $h$ months before or after NGEU and no-NGEU procurement awards, respectively. Note that no-NGEU public procurement bids refer to those different from NGEU-funded bids. The estimation includes firm and sector$\times$time fixed effects, and all standard errors are clustered at the firm level.}
    \end{flushleft}
\end{figure}

Figure \ref{fig:figura9} highlights a significant heterogeneity in the impact of NGEU and no-NGEU bids, particularly on long-term credit. The results indicate that NGEU-funded procurement leads to a 3\% increase in long-term credit one year after the award. In contrast, no-NGEU-funded procurement has a significantly smaller effect, increasing long-term credit by only 0.5\%. 

Regarding short-term credit, the findings suggest that no-NGEU bids primarily translate into short-term financing, leading to a 1\% increase in short-term credit. NGEU bids also have a positive impact on short-term credit; however, the associated estimates exhibit a high degree of uncertainty, making precise quantification less reliable.

In both cases, we find no significant anticipatory effects, indicating that firms do not systematically adjust their borrowing behavior before the award is granted. This suggests that the observed credit dynamics occur as a direct consequence of procurement awards rather than firms preemptively securing financing in anticipation of receiving a contract.

\clearpage

\vspace{0.5cm}

\subsection{From New Lending to changes in the Stock of Credit after NGEU and no-NGEU Awards}

Similarly to the previous analysis, we can translate the impact of receiving NGEU and no-NGEU tenders on new credit in terms of stock by applying the assumption of linear amortization based on the remaining term of the obtained credit. 

\begin{figure}[h!]
    \setlength{\belowcaptionskip}{-5pt} 
    \centering
    \caption{\textbf{Response of New Credit Operations Stock to Public Procurement Bids: NGEU vs no-NGEU}} 
        \vspace{2pt} 
    \includegraphics[width=0.5\textwidth]{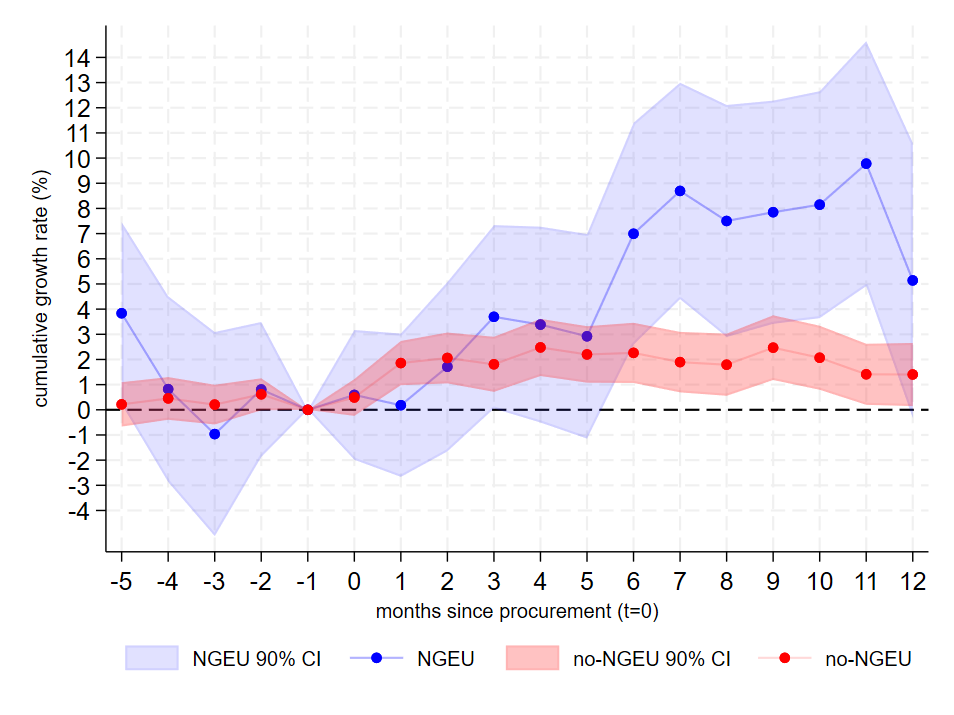}
    \label{fig:figura10}
 \begin{flushleft} 
    \setlength{\parindent}{0pt}
    \setlength{\baselineskip}{8pt} 
        \vspace{-5pt} 
        \scriptsize{
        Notes: the plot displays the estimated coefficients $\beta$ (blue points) and $\gamma$ (red points) from regressions of equation \ref{fig:equation2} for each horizon $h$ relative to 1 month before NGEU and no-NGEU public procurement awards, respectively, as well as their 90\% confidence bands (blue and red shaded areas for NGEU and no-NGEU, respectively). New credit operations stock is calculated by assuming linear credit repayments considering credit term. Thus, it represents an amortization-adjusted new credit operations. Estimated coefficients $\beta$ and $\gamma$ are interpreted as the cumulative growth rate of new credit operations stock $h$ months before or after NGEU and no-NGEU procurement awards, respectively. Note that no-NGEU public procurement bids refer to those different from NGEU-funded bids. The estimation includes firm and sector$\times$time fixed effects, and all standard errors are clustered at the firm level}
    \end{flushleft}
\end{figure}

Figure \ref{fig:figura10} presents the estimation of the $\beta^{h}$ and $\gamma^{h}$ parameters of equation \ref{fig:equation2}, where the dependent variable is the growth rate of the new credit stock. It is observed that one year after receiving NGEU tenders, the new credit stock increases significantly by 5 percentage points, whereas no-NGEU tenders have a milder impact on the stock (in line with the aggregate result). The impact of NGEU tenders on the stock is bounded by existing evidence; specifically, it is slightly lower than the evidence provided by \cite{diGiovanni2022} and higher than that presented by \cite{gabriel2024credit}. This result implies that the impact of NGEU tenders has been in line with the historical impact in Spain.

\vspace{0.5cm}




\section{Robustness checks}

\textbf{Alternative dynamic structure}

As discussed earlier, both specifications included the first lag of the public procurement dummy variables and the dependent variable (new credit, short-term credit, and long-term credit). To assess the robustness of our findings, we test alternative dynamic structures and lag specifications. Specifically, we extend the number of lags to three and six to determine whether additional lags of the main regressor and the dependent variable influence the estimated impact of public procurement on new credit operations over time.

Figure \ref{fig:figura13} in the Appendix presents the results of estimating equation \ref{fig:equation1} (the effect of all public procurement bids on new credit) using three and six lags. The results remain consistent with those previously reported, indicating that the estimated effects are not sensitive to lag length. Similarly, Figure \ref{fig:figura14} in the Appendix shows the results of equation \ref{fig:equation2} (which differentiates between NGEU and no-NGEU procurement effects) under both alternative lag structures. The findings remain unchanged, confirming that the estimated effects of procurement on new credit are robust to different lag parametrizations.

\textbf{Time sample selection}

Since our database includes the onset of the pandemic, it is possible that the impact of no-NGEU tenders on new credit may have changed before and after the pandemic. One potential explanation is that the introduction of NGEU funds may have led to a substitution effect, where tenders that were previously financed through other sources were instead funded by NGEU, amplifying their impact while reducing the relative effect of no-NGEU tenders.\footnote{Before the introduction of NGEU funds, certain projects would have been financed by alternative sources. With the availability of NGEU funds, some of these tenders may have shifted to the new financing mechanism, possibly affecting the observed impact of no-NGEU tenders.}

To test whether the impact of no-NGEU public tenders on new credit changed due to the introduction of NGEU funds, we re-estimate equation \ref{fig:equation2}, restricting the sample to July 2020 onward, when European Council approved the implementation of the extraordinary instrument of temporary recovery; NGEU funds. Figure \ref{fig:figura15} in the Appendix presents the results, which remain consistent with those from the full sample. This indicates that the effect of no-NGEU public procurement on new credit remains stable across both time periods, suggesting that the introduction of NGEU funds did not significantly alter the credit dynamics of no-NGEU tenders.

\textbf{Expert procurement firms}

To better understand the relationship between public tenders and new credit, we examine whether the observed effects are primarily driven by firms referred to as "experts"—those that have received both no-NGEU and NGEU tenders. These firms possess significant experience in the public procurement process, often leveraging internal specialized departments or external consulting firms to prepare and submit tender documents for public competitions.

To assess this, we restrict the sample to companies that have been awarded both types of tenders at some point.\footnote{Fraction of firms that have received both no-NGEU and NGEU tenders.} We then re-estimate equation \ref{fig:equation2}, with the results presented in Figure \ref{fig:figura16} in the Appendix. The findings remain consistent with those from the full sample of NGEU funded firms, indicating that the effects of NGEU public procurement on new credit are largely driven by this group of experienced firms. Furthermore, it is noteworthy that the impulse on new credit after no-NGEU awards remains positive, albeit not significantly. This implies that there is greater uncertainty associated with it, indicating that expert firms react significantly in terms of new credit following the receipt of NGEU tenders, but not in the case of no-NGEU tenders. Behind this phenomenon, substitution effects between the two types of tenders may be at play.

\section{Conclusions}
\label{sec:conclusion}

This study provides robust empirical evidence that public procurement awards significantly enhance firm-level new lending, generating a cumulative increase of 0.75\% in new credit operations within a year. The effect becomes statistically significant six months after contract allocation and persists throughout the first year, reinforcing procurement’s critical role in firm financing.

A key finding is that Next Generation EU (NGEU)-funded contracts elicit a stronger credit response than traditional procurement, highlighting the effectiveness of targeted European recovery funds in enhancing liquidity. Specifically, NGEU-funded procurement leads to a 3\% increase in long-term credit one year after the award, whereas no-NGEU-funded procurement has a more limited effect, increasing long-term credit by only 0.5\% over the same period. In contrast, no-NGEU bids primarily stimulate short-term financing, resulting in a 1\% increase in short-term credit.

To compare our new lending findings with the equivalent results in prior research, we adjusted our data for amortization to measure changes in credit stock. This approach yields a dynamic elasticity of 1.5 percentage points one year after procurement allocation. Although this figure is lower than the 5.5 percentage points for Spain reported by \cite{diGiovanni2022} and the 3 percentage points for Portugal found by \cite{gabriel2024credit}, these differences likely stem from variations in sample composition, firm characteristics, and procurement contract structures. Notably, our 90\% confidence interval’s upper bound is 3.0 percentage points, which closely matches  the ones for Portugal by \cite{gabriel2024credit}. Moreover, when we focus on NGEU-funded procurement alone, the dynamic elasticity rises to 5\%, slightly lower than the historical 6\% documented by \cite{diGiovanni2022}. Furthermore, our analysis spans both no-NGEU and NGEU periods, whereas \cite{diGiovanni2022} examine a strictly pre-NGEU timeframe, potentially contributing to the observed divergence. Despite these discrepancies, we find no evidence of anticipatory effects on credit stock, consistent with \cite{gabriel2024credit}.

The impact of procurement on new lending varies across firm size, industry, and value chain position. Smaller firms exhibit the strongest response, with credit elasticity reaching 1.25\% one year after contract allocation, compared to 0.5\% for larger firms. Firms in government-dependent sectors, such as construction and manufacturing, also experience the largest credit boosts. Additionally, procurement-induced financing is predominantly used for working capital rather than long-term investment, as short-term credit expands more than long-term credit.

Further analysis shows that while procurement enhances liquidity across firms, the magnitude and timing of credit expansion exhibit significant heterogeneity based on firm characteristics. The findings suggest that smaller, more financially constrained firms benefit disproportionately from procurement contracts, supporting the view that public procurement can act as a financial catalyst for firms otherwise facing borrowing limitations. Similarly, government-dependent industries exhibit a pronounced response, indicating that procurement awards play a stabilizing role in sectors where public spending constitutes a significant share of demand. 

Distinct patterns in credit responses emerge based on a firm’s position in the value chain. Downstream firms (closer to final demand) tend to experience an immediate surge in credit availability following procurement awards, whereas upstream firms (earlier in the supply chain) exhibit a delayed but more persistent expansion. This dynamic suggest that procurement effects could propagate heterogeneously through the production network: firms nearer to end-users benefit sooner, while those further up the supply chain face a lagged yet enduring impact. However, the difference in timing is statistically significant only when comparing the immediate expansions in downstream firms to the delayed expansions in upstream firms. Nevertheless, the magnitude of these idiosyncratic procurement shocks appears weaker than the effects documented for a global monetary policy shock by \cite{buda2023}.

This study employs a unique dataset that integrates public procurement records from the Spanish Ministry of Finance with high-frequency new lending data from BBVA, enabling an unprecedented analysis of firm-level credit dynamics. The dataset includes over 2,000 firms receiving NGEU funding and more than 17,000 firms engaged in no-NGEU procurement, covering nearly 240,000 credit transactions and around 120,000 procurement awards. By incorporating detailed, high-frequency financial data, this study provides a more granular assessment of procurement’s role in credit expansion, thereby yielding new insights into firm-level financial responses

Beyond its direct contributions to the literature on public procurement and firm financing, this study enhances the broader understanding of government spending’s impact on economic growth. It also contributes to the expanding field of high-frequency economic analysis, demonstrating the value of transaction-level data in capturing real-time shifts in firm behavior and credit markets.

From a policy standpoint, the use of high-frequency, granular data is pivotal in refining economic interventions—particularly procurement strategies. By closely tracking firm-level credit responses, policymakers can more accurately identify which businesses or sectors are most sensitive or vulnerable, thereby informing more targeted and impactful contract designs. Such fine-grained insights also clarify how liquidity effects ripple through supply chains, helping to develop adaptive, evidence-based procurement policies that enhance overall economic resilience.

\newpage
\section{Appendix}

\subsection{Data descriptive statistics}

In this section, we describe the main characteristics of the firms under analysis, distinguishing four groups: All Procurement, the NGEU program, no-NGEU funds, and Experts (defined as firms that received both NGEU and no‐NGEU funds). Table \ref{tab:Tabla2} presents descriptive statistics for key firm characteristics, including firm age, number of employees, turnover, new credit (disaggregated into new short-term and long-term credit), and new credit term. For each group, the table reports the distribution of these variables, providing the mean, coefficient of variation, the 25th, 50th, and 75th percentiles, as well as the minimum and maximum observed values. This offers a preliminary overview of central tendencies and variability within each group.

\begin{table}[h]
    \centering
    \caption{\textbf{Descriptive statistics: the Database}}
    \vspace{0.3cm}
    \label{tab:Tabla2}
    \renewcommand{\arraystretch}{1.3}
    \resizebox{\textwidth}{!}{
        \begin{tabular}{|c|c|c|c|c|c|c|c|c|}\toprule
             \multicolumn{2}{|c|}{} & Mean & CV & p25 & p50 & p75 & Min & Max\\ \midrule 
             \multirow{7}{*}{\textbf{All Procurement}} & \textbf{Age} & 24.5 & 0.6 & 13.0 & 23.0 & 33.0 & 1.0 & 123.0\\
             & \textbf{Employees} & 113.4 & 10.5 & 7.0 & 16.0 & 45,0 & 0.0 & 104542.0 \\
             & \textbf{Turnover} & 21474346.1 & 13.2 & 755159.2 & 2164792.3 & 7408882.0 & 0.0 & 27820314624.0 \\
             & \textbf{New credit} & 377820.9 & 8.2 & 13868.6 & 47470.8 & 152694.5 & 0.0 & 250001435.9\\
             & \textbf{New short-term credit} & 320318.8 & 9.0 & 10888.2 & 37102.0 & 122406.3 & 0.5 & 215441068.4 \\
             & \textbf{New long-term credit} & 629506.0 & 6.3 & 40361.4 & 107300.0 & 301572.0 & 0.0 & 250001435.9 \\
             & \textbf{New credit term} & 270.0 & 9.3 & 36.5 & 84.0 & 111.0 & 0.0 & 738396.0\\
             \hline
             \multirow{7}{*}{\textbf{NGEU}} & \textbf{Age} & 24.6 & 0.6 & 13.0 & 23.0 & 33.0 & 1.0 & 123.0\\
             & \textbf{Employees} & 163.2 & 5.4 & 11.0 & 28.0 & 80.0 & 1.0 & 28531.0\\
             & \textbf{Turnover} & 32430455.1 & 5.0 & 1464494.9 & 4573926.0 & 15806635.0 & 100.0 & 4887763968.0\\
             & \textbf{New credit} & 432623.2 & 7.1 & 24000.0 & 76710.9 & 240992.6 & 1.0 & 250001435.9\\
             & \textbf{New short-term credit} & 338909.8 & 5.1 & 20000.0 & 63000.0 & 200000.0 & 1.0 & 85000000.0\\
             & \textbf{New long-term credit} & 847284.4 & 7.3 & 55277.4 & 150000.0 & 400876.7 & 1.0 & 250001435.9\\
             & \textbf{New credit term} & 223.8 & 5.5 & 41.0 & 86.0 & 120.0 & 0.0 & 191710.0\\
             \hline
             \multirow{7}{*}{\textbf{no-NGEU}} & \textbf{Age} & 24.5 & 0.6 & 13.0 & 23.0 & 33.0 & 1.0 & 123.0\\
             & \textbf{Employees} & 114.1 & 10.5 & 7.0 & 16.0 & 45.0 & 0.0 & 104542.0\\
             & \textbf{Turnover} & 21555417.5 & 13.2 & 755534.3 & 2165177.3 & 7399208.0 & 0.0 & 27820314624.0\\
             & \textbf{New credit} & 379053.9 & 8.2 & 13797.3 & 47157.0 & 152097.8 & 0.0 & 250001435.9\\
             & \textbf{New short-term credit} & 321776.5 & 9.0 & 10823.2 & 36941.3 & 122077.5 & 0.5 & 215441068.4\\
             & \textbf{New long-term credit} & 629648.4 & 6.3 & 40118.7 & 106000.0 & 300000.0 & 0.0 & 250001435.9\\
             & \textbf{New credit term} & 270.7 & 9.3 & 36.6 & 84.0 & 111.0 & 0.0 & 738396.0\\
             \hline
             \multirow{7}{*}{\textbf{Experts}} & \textbf{Age} & 24.9 & 0.6 & 14.0 & 23.0 & 33.0 & 1.0 & 123.0\\
             & \textbf{Employees} & 171.6 & 5.3 & 11.0 & 30.0 & 83.0 & 1.0 & 28531.0\\
             & \textbf{Turnover} & 33836111.3 & 4.9 & 1545982.6 & 4729065.5 & 16788286.0 & 3000.0 & 4887763968.0\\
             & \textbf{New credit} & 444809.6 & 7.1 & 23951.3 & 77142.4 & 243210.7 & 1.0 & 250001435.9\\
             & \textbf{New short-term credit} & 350437.7 & 5.1 & 20000.0 & 63310.5 & 200000.0 & 1.0 & 85000000.0\\
             & \textbf{New long-term credit} & 862997.9 & 7.4 & 54538.3 & 150000.0 & 400000.0 & 1.0 & 250001435.9\\
             & \textbf{New credit term} & 225.4 & 5.6 & 41.0 & 86.9 & 120.0 & 0.0 & 191710.0\\
             \bottomrule
    \end{tabular}
    }
\end{table}

Overall, firms receiving a procurement contract have an average age of 24.5 years, with ages ranging from 1 to 123 years. On average, these firms employ 214,744 individuals; however, a maximum of 2,057,264 employees and a coefficient of variation of 4.3 indicate substantial heterogeneity in firm size. Similarly, turnover displays a wide range (minimum of 1, maximum of 9.7 billion), with a mean of 37.8 million and a coefficient of variation of 8.9, suggesting that several large outliers skew the average.

Within the NGEU subset, the average firm age is comparable to that of the overall sample (mean: 24.4 years), whereas the average number of employees is 215,535, reflecting a skewed distribution driven by several large firms. The no‐NGEU group exhibits a similar pattern in age (mean: 24.5 years) and number of employees (mean: 215,351.5), indicating that these subgroups are broadly comparable in terms of firm age and size. In fact, 92\% of the firms that received NGEU funds also received no‐NGEU funds, further underscoring this similarity.

The Experts group, that is, firms that received both NGEU and no‐NGEU funds, is smaller in number but is characterized by notably larger mean values for both employees (338,361.1) and turnover (129.8 million) compared to the other groups. Moreover, the higher median values for employees (67,201) and turnover (9.1 million) indicate that expert firms tend to be larger.

Regarding financial indicators, new credit amounts are relatively modest in the All Procurement group (mean: 27.0) compared to the Experts group (mean: 225.4). A similar pattern is observed for new short-term and long-term credit, with the Experts group consistently exhibiting higher mean values. This suggests that expert firms may secure larger credit lines or more significant financing packages. Correspondingly, the NGEU and no‐NGEU groups demonstrate similar levels of new credit usage (means of 8.2 and 7.7, respectively), although NGEU firms display slightly higher dispersion (coefficient of variation: 2.9 versus 2.6). Finally, new credit term (measured in months) shows limited variation across groups, indicating that the maturities of credit agreements do not differ substantially among these firms.

In summary, the descriptive statistics highlight substantial heterogeneity in firm size (number of employees) and turnover across all groups, with the Experts group consistently exhibiting the highest averages. Although age distributions are broadly similar, financing patterns—particularly the magnitude of new credit obtained—vary more noticeably across groups. These findings provide a baseline understanding of the sample characteristics, which provides knowledge of the analysis and the interpretation of our model’s results.

\subsection{Additional quantitative analysis}

\label{sec:appendix}


\vspace{0.5cm}
\begin{figure}[h!]
    \centering
    \caption{\textbf{Response of New Credit Operations to Public Procurement Bids without COVID-related credits: New Credit and New Credit Stock}} 
    \vspace{2pt} 
    \label{fig:figura11}
    \begin{subfigure}[b]{0.45\textwidth} 
        \centering
        \includegraphics[width=\textwidth]{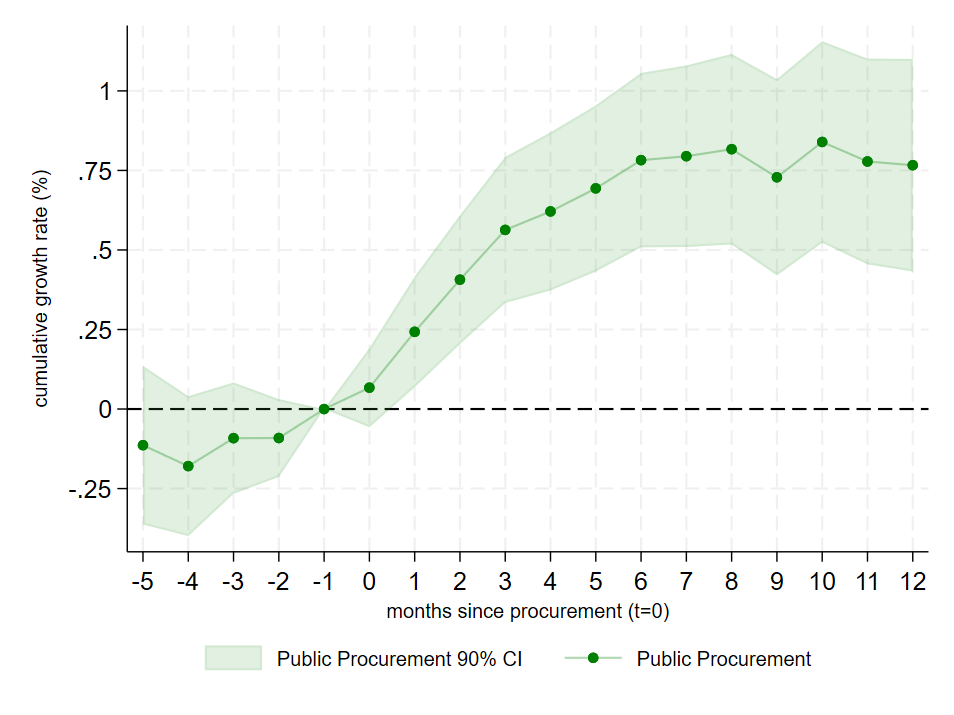} 
        \caption{\textbf{New credit}} 
        \label{fig:graph1b}
    \end{subfigure}
    \hspace{-5pt} 
    \begin{subfigure}[b]{0.45\textwidth} 
        \centering
        \includegraphics[width=\textwidth]{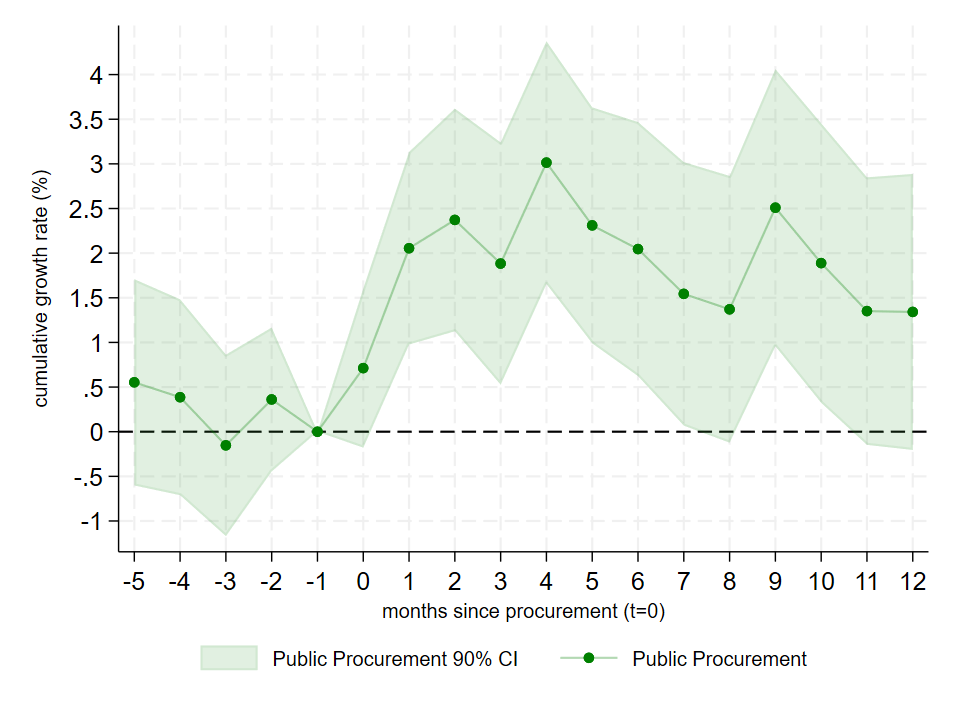} 
        \caption{\textbf{New credit stock}} 
        \label{fig:graph2b}
    \end{subfigure}
    
    \vspace{-5pt} 
    \begin{flushleft} 
    \scriptsize
    \setlength{\parindent}{0pt}
    \setlength{\baselineskip}{8pt} 
    
        \vspace{-5pt} 
        \scriptsize{
        Notes: The plots display the estimated coefficient $\beta$ (green points) from regressions of equation \ref{fig:equation1} for each horizon $h$ relative to 1 month before public procurement awards, as well as its 90\% confidence bands (green shaded area). Panel (a) shows the results for the case of the dependent variable being firm new credit after subtracting COVID-related credits, and panel (b) is similar to (a) but after transforming to credit stock by considering linear repayments. Estimated coefficient $\beta$ is interpreted as the cumulative growth rate of new credit operations (a) and new credit operations stock (b) $h$ months before or after procurement awards. The estimation includes firm and sector$\times$time fixed effects, and all standard errors are clustered at the firm level}
    \end{flushleft}
\end{figure}

\vspace{0.5cm}
\begin{figure}[h!]
    \centering
    \caption{\textbf{Response of New Credit Operations to Public Procurement Bids by Upstreamness: Downstream vs Upstream}} 
    \vspace{2pt} 
    \label{fig:figura12}
    \begin{subfigure}[b]{0.45\textwidth} 
        \centering
        \includegraphics[width=\textwidth]{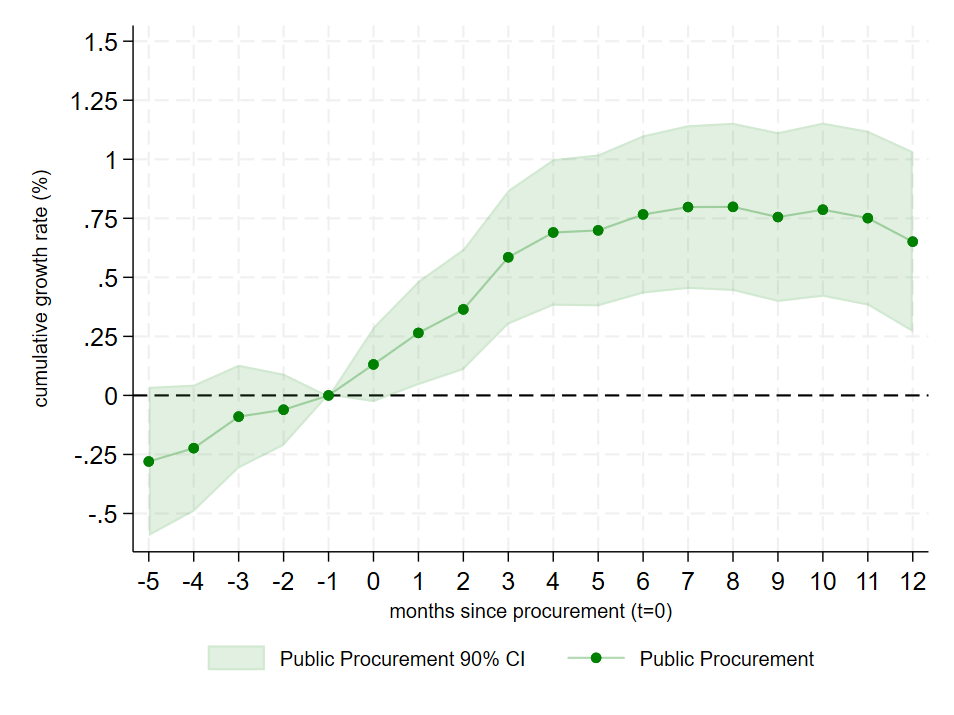} 
        \caption{\textbf{Downstrem}} 
    \end{subfigure}
    \hspace{-5pt} 
    \begin{subfigure}[b]{0.45\textwidth} 
        \centering
        \includegraphics[width=\textwidth]{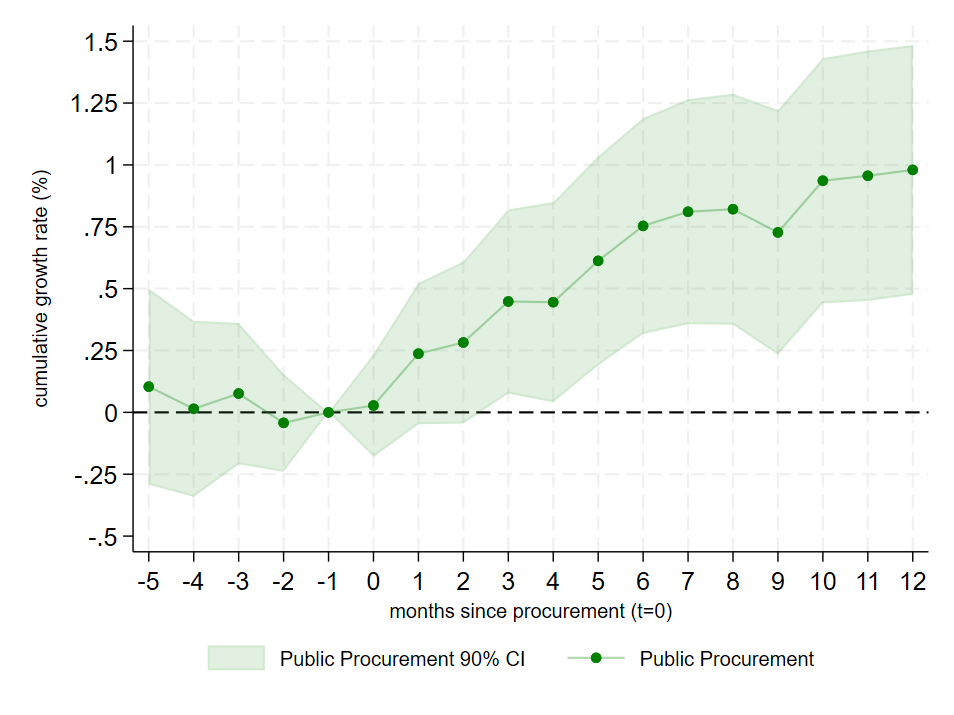} 
        \caption{\textbf{Upstream}} 
        \label{fig:graph2_unique1}
    \end{subfigure}
    
    \begin{flushleft} 
    \scriptsize
    \setlength{\parindent}{0pt}
    \setlength{\baselineskip}{8pt} 
    
        \vspace{-5pt} 
        \scriptsize{
         Notes: The plots display the estimated coefficient $\beta$ from regressions of equation \ref{fig:equation1} for each horizon $h$ relative to 1 month before public procurement awards, as well as its 90\% confidence bands (shaded areas), for the case of downstream firms (panel (a)) and upstream (panel (b)). Panel (a) includes firms classified as downstream, or relatively close to the consumer, if the metric is equal to or less than 2.2, and panel (b) firms classified as upstream, or distant from the final consumer (upstreamness indicator greater than 2.2). The categorization is based on \cite{antras2012measuring}, and further developed for the case of Spain by \cite{buda2023short}. The estimated coefficient $\beta$ is interpreted as the cumulative growth rate of new credit operations (by upstreamness degree) $h$ months before or after procurement awards. The estimation includes firm and sector$\times$time fixed effects, and all standard errors are clustered at the firm level.}
    \end{flushleft}
\end{figure}



\begin{figure}[h!]
    \centering
    \caption{\textbf{Response of New Credit Operations to Public Procurement Bids: Different Lag Parametrizations}} 
    \vspace{2pt} 
    \label{fig:figura13}
    \begin{subfigure}[b]{0.45\textwidth} 
        \centering
        \includegraphics[width=\textwidth]{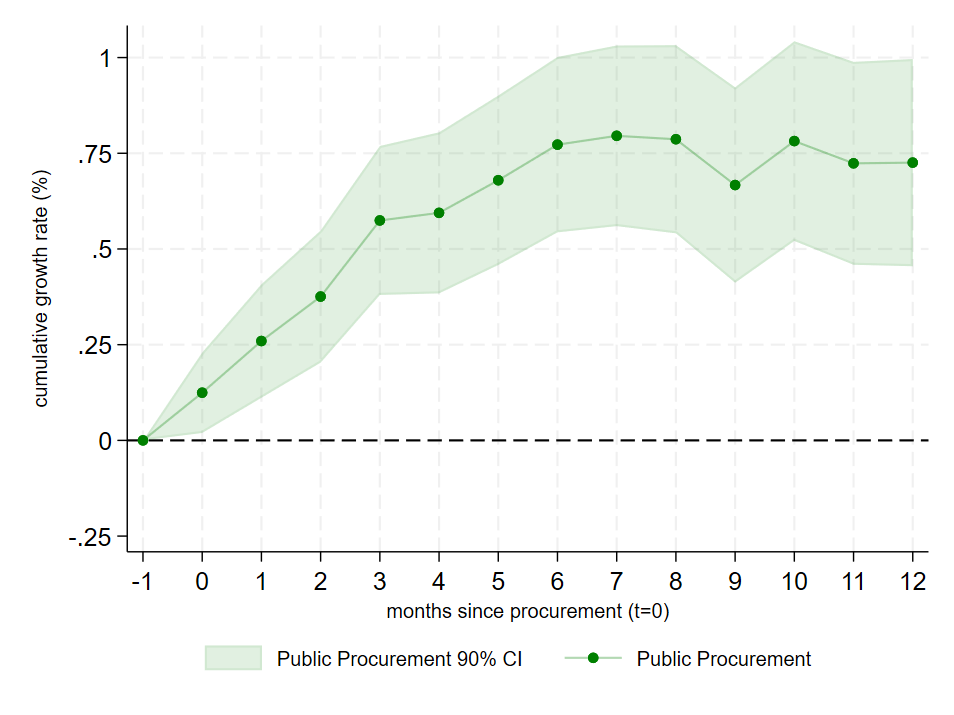} 
        \caption{\textbf{3 lags}} 
        \label{fig:graph1_2}
    \end{subfigure}
    \hspace{-5pt} 
    \begin{subfigure}[b]{0.45\textwidth} 
        \centering
        \includegraphics[width=\textwidth]{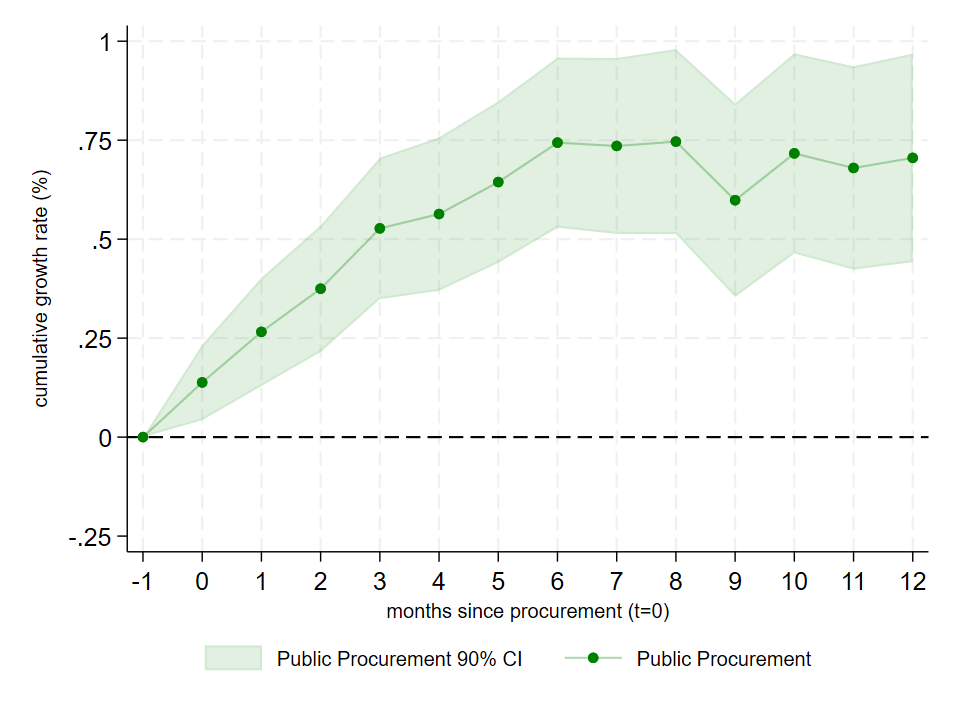} 
        \caption{\textbf{6 lags}} 
        \label{fig:graph2}
    \end{subfigure}
    
    \vspace{-5pt} 
    \begin{flushleft} 
    \scriptsize
    \setlength{\parindent}{0pt}
    \setlength{\baselineskip}{8pt} 
    
        \vspace{-5pt} 
        \scriptsize{
        Notes: The plots display the estimated coefficient $\beta$ (green points) from regressions of equation \ref{fig:equation1} for each horizon $h$ relative to 1 month before public procurement awards, as well as its 90\% confidence bands (green shaded area). Panel (a) shows the results for the case of controlling for 3 lags of the public procurement dummy and dependent variable, and panel (b) is similar to (a) but controlling for 6 lags. Estimated coefficient $\beta$ is interpreted as the cumulative growth rate of new credit operations $h$ months after procurement awards. Anticipatory exploration is excluded given the dynamic conflict of lags with the second element of the dependent variable. The estimation includes firm and sector$\times$time fixed effects, and all standard errors are clustered at the firm level}
    \end{flushleft}
\end{figure}


\begin{figure}[h!]
    \centering
    \caption{\textbf{Response of New Credit Operations to Public Procurement Bids NGEU vs no-NGEU: Different Lag Parametrizations}} 
    \vspace{2pt} 
    \label{fig:figura14}
    \begin{subfigure}[b]{0.45\textwidth} 
        \centering
        \includegraphics[width=\textwidth]{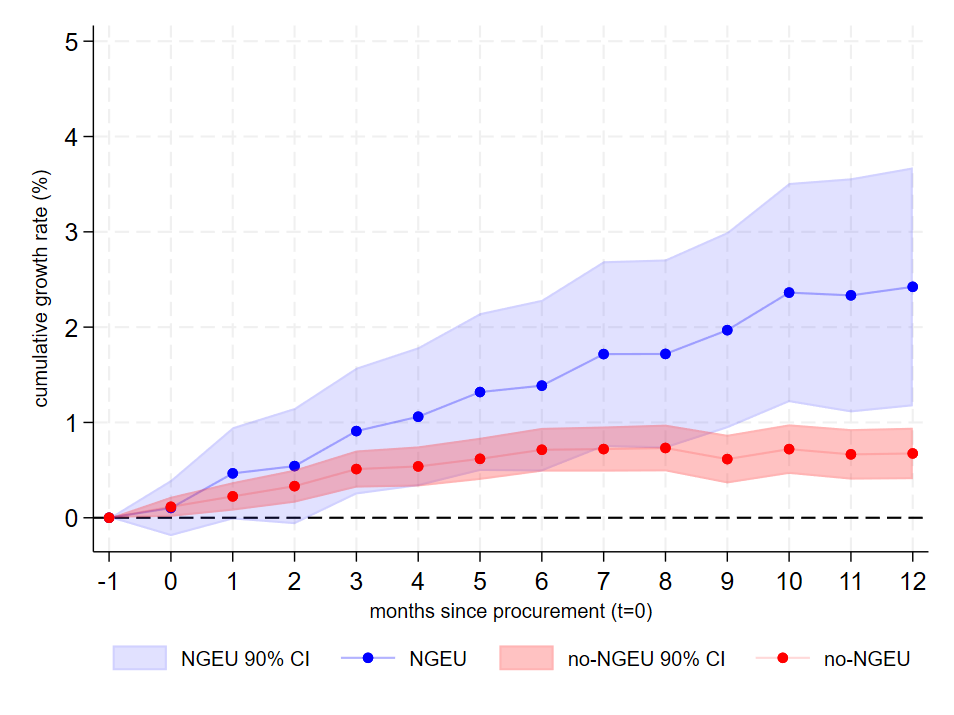} 
        \caption{\textbf{3 lags}} 
        \label{fig:graph1}
    \end{subfigure}
    \hspace{-5pt} 
    \begin{subfigure}[b]{0.45\textwidth} 
        \centering
        \includegraphics[width=\textwidth]{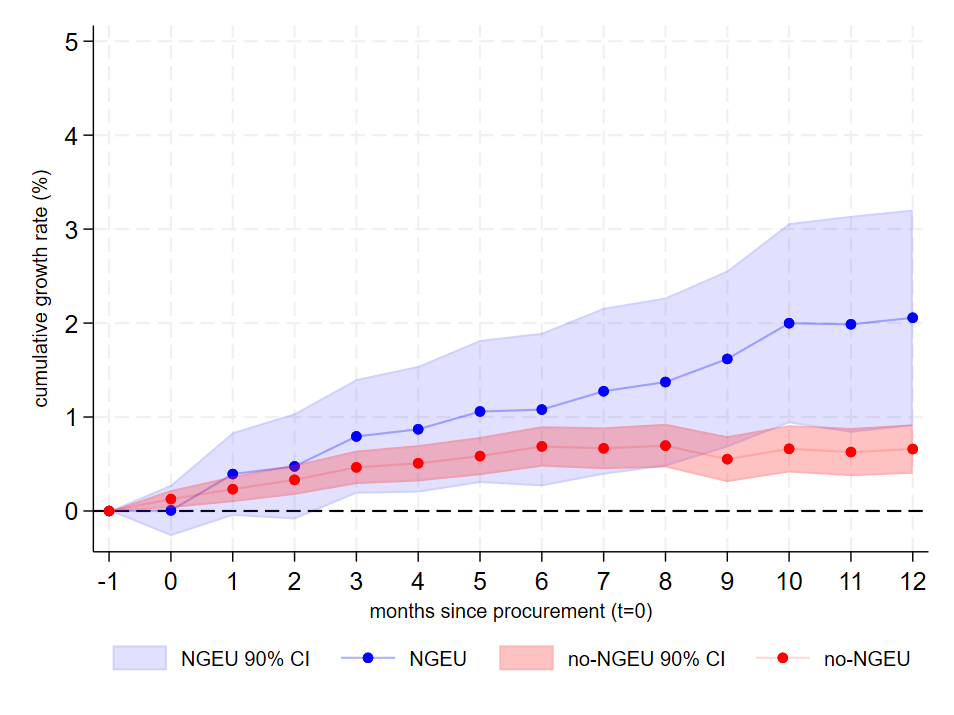} 
        \caption{\textbf{6 lags}} 
        \label{fig:graph2}
    \end{subfigure}
    
    \vspace{-5pt} 
    \begin{flushleft} 
    \scriptsize
    \setlength{\parindent}{0pt}
    \setlength{\baselineskip}{8pt} 
    
        \vspace{-5pt} 
        \scriptsize{
         Notes: the plot displays the estimated coefficients $\beta$ (blue points) and $\gamma$ (red points) from regressions of equation \ref{fig:equation2} for each horizon $h$ relative to 1 month before NGEU and no-NGEU public procurement awards, respectively, as well as their 90\% confidence bands (blue and red shaded areas for NGEU and no-NGEU, respectively). Panel (a) shows the results for the case of controlling for 3 lags of the public procurement dummies and dependent variable, and panel (b) is similar to (a) but controlling for 6 lags. Estimated coefficients $\beta$ and $\gamma$ are interpreted as the cumulative growth rate of new credit operations $h$ months before or after NGEU and no-NGEU procurement awards, respectively. Note that no-NGEU public procurement bids refer to those different from NGEU-funded bids. The estimation includes firm and sector$\times$time fixed effects, and all standard errors are clustered at the firm level.}
    \end{flushleft}
\end{figure}


\begin{figure}[h!]
    \setlength{\belowcaptionskip}{-5pt} 
    \centering
    \caption{\textbf{Response of New Credit Operations to Public Procurement Bids: NGEU vs no-NGEU (since NGEU EU Council Approval)}} 
        \vspace{2pt} 
    \includegraphics[width=0.5\textwidth]{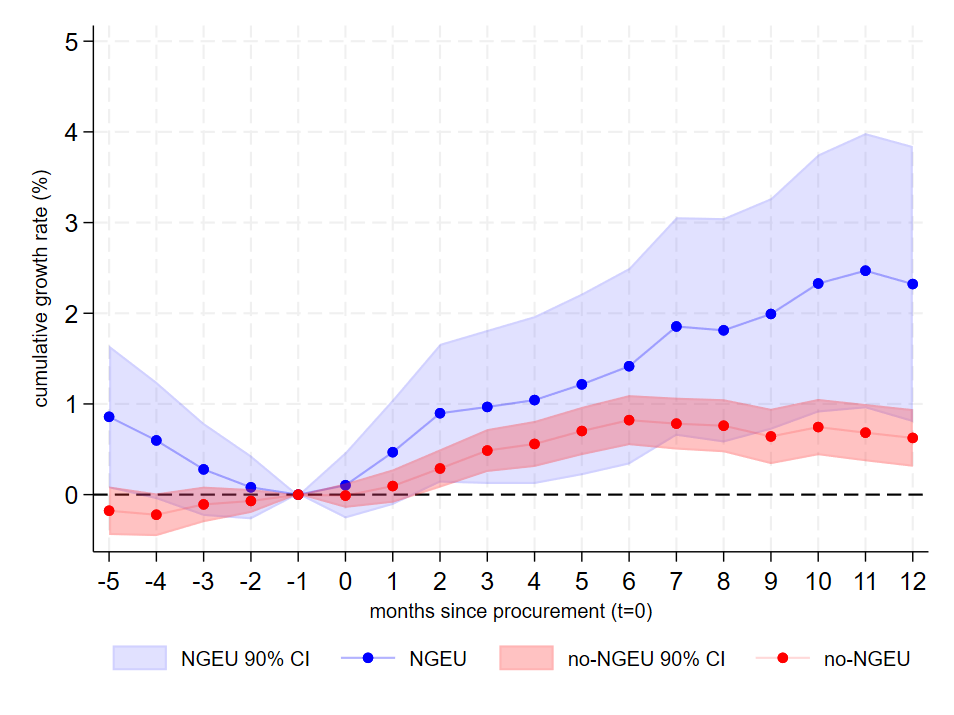}
    \label{fig:figura15}

    \begin{flushleft} 
    \scriptsize
    \setlength{\parindent}{0pt}
    \setlength{\baselineskip}{8pt} 
    
        \vspace{-5pt} 
        \scriptsize{
        Notes: the plot displays the estimated coefficients $\beta$ (blue points) and $\gamma$ (red points) from regressions of equation \ref{fig:equation2} for each horizon $h$ relative to 1 month before NGEU and no-NGEU public procurement awards, respectively, as well as their 90\% confidence bands (blue and red shaded areas for NGEU and no-NGEU, respectively). The time sample has been restricted to begin at the time when the EU Council approved the implementation of the NGEU program (July 2020). Estimated coefficients $\beta$ and $\gamma$ are interpreted as the cumulative growth rate of new credit operations $h$ months before or after NGEU and no-NGEU procurement awards, respectively. Note that no-NGEU public procurement bids refer to those different from NGEU-funded bids. The estimation includes firm and sector$\times$time fixed effects, and all standard errors are clustered at the firm level.}
    \end{flushleft}
\end{figure}

\begin{figure}[h!]
    \setlength{\belowcaptionskip}{-5pt} 
    \centering
    \caption{\textbf{Response of New Credit Operations to Public Procurement Bids: NGEU vs no-NGEU (Expert Firms)}} 
        \vspace{2pt} 
    \includegraphics[width=0.5\textwidth]{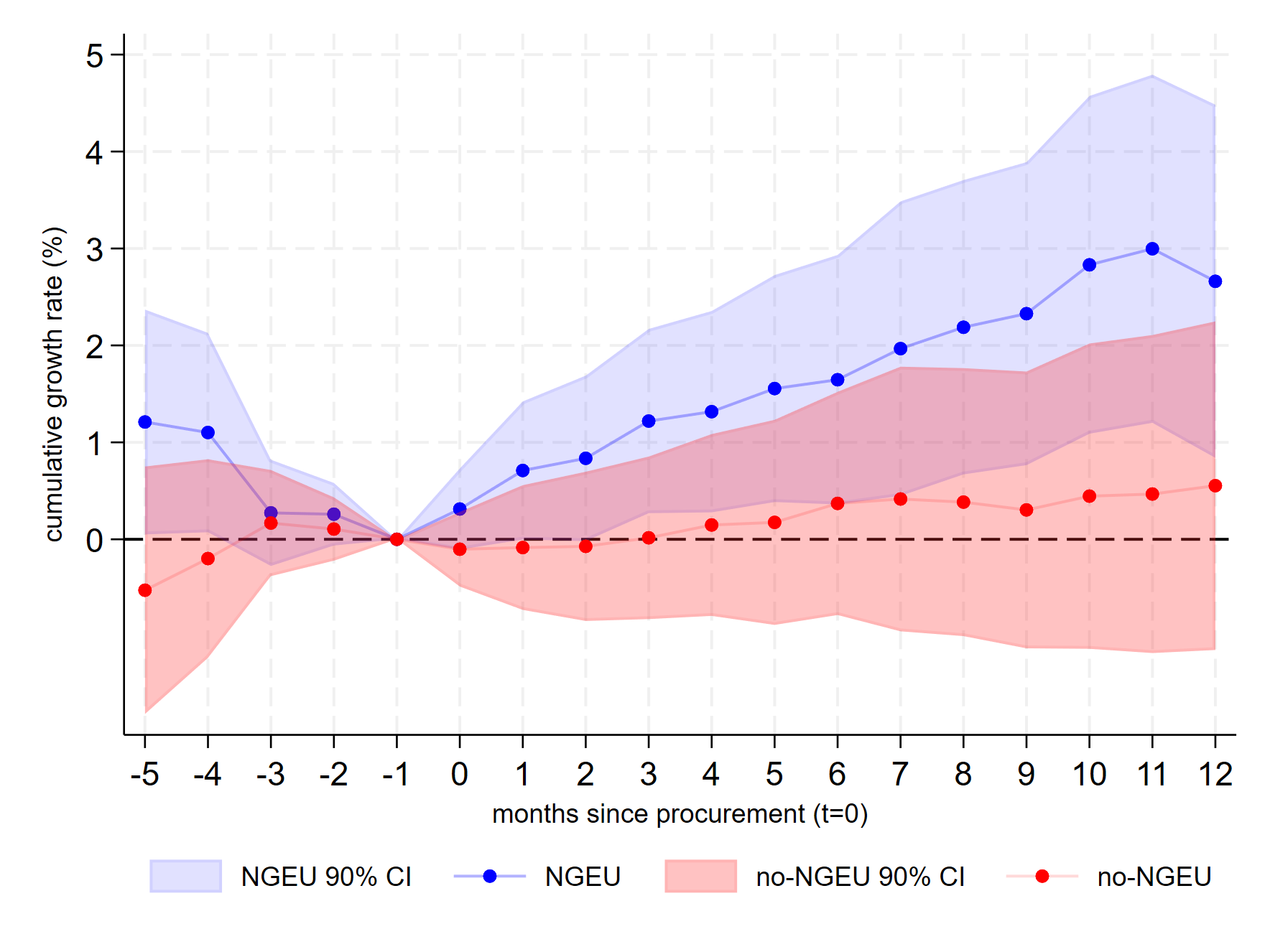}
    \label{fig:figura16}

    \begin{flushleft} 
    \scriptsize
    \setlength{\parindent}{0pt}
    \setlength{\baselineskip}{8pt} 
    
        \vspace{-5pt} 
        \scriptsize{
        Notes: the plot displays the estimated coefficients $\beta$ (blue points) and $\gamma$ (red points) from regressions of equation \ref{fig:equation2} for each horizon $h$ relative to 1 month before NGEU and no-NGEU public procurement awards, respectively, as well as their 90\% confidence bands (blue and red shaded areas for NGEU and no-NGEU, respectively). The has been restricted to those firms, denominated as experts, that have been awarded NGEU and no-NGEU bids at some point in the time sample. Estimated coefficients $\beta$ and $\gamma$ are interpreted as the cumulative growth rate of new credit operations $h$ months before or after NGEU and no-NGEU procurement awards, respectively. Note that no-NGEU public procurement bids refer to those different from NGEU-funded bids. The estimation includes firm and sector$\times$time fixed effects, and all standard errors are clustered at the firm level.}
    \end{flushleft}
\end{figure}

\clearpage

\subsection{Upstream vs. Downstream Sectoral Classification }\label{subsec:further_upstream}

In this section, we describe the followed methodology to bridge the sector classification used by the Spanish Tax Authority to compile their sales data with the 2015 Spanish INE Input-Output sector classification and the computation of the upstreamness indicator and the sectoral upstream classification for the sales data according to \cite{buda2023}. The authors present a mapping between the Spanish Tax Authority and the 2015 Spanish INE Input-Output sector classification. Out of the 64 sectors listed in the INE IO table, they match 43 to the sales sector classification. Ultimately, they link 20 sales sectors to the 43 sectors in the IO table, for which they calculate an upstreamness indicator following the upstreamness metric proposed by \cite{antras2012}, designed  to accurately assess an industry position within the global production network. The upstreamness metric offers an insightful perspective on the relative distance of an industry from the final consumption phase, reflecting their involvement in early or intermediate stages in the supply chain. These authors show that this metric can be derived taking two distinct approaches, both anchored in input-output analysis, and both of them produce an upstreamness measure that is always at least one for every industry. Higher values indicate a greater degree of upstreamness for that particular industry.

Table \ref{tab:Tabla3} reports the upstreamness indicator for the industries we can monitor. The first and second column reports the industry classification at NACE code 2 digits, respectively. The third column, the “Upstreamness” values presented in the table range
from 1.01 for residential care services to 3.84 for mining support services activities, demonstrating
the variation in industries positions within the supply chain.

\begin{table}[h]
    \centering
    \caption{\textbf{Descriptive statistics: the Database}}
    \vspace{0.1cm}
    \label{tab:Tabla3}
    \renewcommand{\arraystretch}{0.97}
    \resizebox{\textwidth}{!}{
        \begin{tabular}{|p{20cm}|c|c|}\toprule
            Description & CNAE-2-digit & Up\\ \midrule
            Residential care activities & 87 & 1,01 \\
            Social work activities without accommodation & 88 & 1,01 \\
            Education & 85 & 1,12 \\
            Human health activities & 86 & 1,14 \\
            Public administration and defence; compulsory social security & 84 & 1,16 \\
            Accommodation & 55 & 1,18 \\
            Food and beverage service activities & 56 & 1,18 \\
            Creative, arts and entertainment activities & 90 & 1,42 \\
            Libraries, archives, museums and other cultural activities & 91 & 1,42 \\
            Gambling and betting activities & 92 & 1,42 \\
            Manufacture of basic pharmaceutical products and pharmaceutical preparations & 21 & 1,55 \\
            Sports activities and amusement and recreation activities & 93 & 1,57 \\
            Real estate activities & 68 & 1,58 \\
            Activities of households as employers of domestic personnel & 97 & 1,70 \\
            Undifferentiated goods- and services-producing activities of private households for own use & 98 & 1,70 \\
            Activities of membership organisations & 94 & 1,76 \\
            Wholesale and retail trade and repair of motor vehicles and motorcycles & 45 & 1,77 \\
            Wholesale trade, except of motor vehicles and motorcycles & 46 & 1,77 \\
            Retail trade, except of motor vehicles and motorcycles & 47 & 1,77 \\
            Construction of buildings & 41 & 1,83 \\
            Civil engineering & 42 & 1,83 \\
            Specialised construction activities & 43 & 1,83 \\
            Insurance, reinsurance and pension funding, except compulsory social security & 65 & 1,89 \\
            Activities auxiliary to financial services and insurance activities & 66 & 1,89 \\
            Repair of computers and personal and household goods & 95 & 1,89 \\
            Other personal service activities & 96 & 1,89 \\
            Manufacture of leather and related products & 15 & 1,98 \\
            Manufacture of motor vehicles, trailers and semi-trailers & 29 & 2.00 \\
            Manufacture of other transport equipment & 30 & 2.00 \\
            Manufacture of textiles & 13 & 2,02 \\
            Manufacture of wearing apparel & 14 & 2,02 \\
            Manufacture of computer, electronic and optical products & 26 & 2,09 \\
            Manufacture of electrical equipment & 27 & 2,09 \\
            Repair and installation of machinery and equipment & 33 & 2,17 \\
            Water collection, treatment and supply & 36 & 2,18 \\
            Waste collection, treatment and disposal activities; materials recovery & 38 & 2,18 \\
            Remediation activities and other waste management services & 39 & 2,18 \\
            Manufacture of wood and of products of wood and cork, except furniture; manufacture of articles of straw and plaiting materials & 16 & 2,19 \\
            Publishing activities & 58 & 2,20 \\
            Motion picture, video and television programme production, sound recording and music publishing activities & 59 & 2,20 \\
            Programming and broadcasting activities & 60 & 2,20 \\
            Telecommunications & 61 & 2,20 \\
            Computer programming, consultancy and related activities & 62 & 2,20 \\
            Information service activities & 63 & 2,20 \\
            Manufacture of food products & 10 & 2,27 \\
            Manufacture of beverages & 11 & 2,27 \\
            Manufacture of tobacco products & 12 & 2,27 \\
            Manufacture of machinery and equipment n.e.c. & 28 & 2,30 \\
            Crop and animal production, hunting and related service activities & 1 & 2,42 \\
            Forestry and logging & 2 & 2,42 \\
            Fishing and aquaculture & 3 & 2,42 \\
            Manufacture of furniture & 31 & 2,48 \\
            Postal and courier activities & 53 & 2,52 \\
            Financial service activities, except insurance and pension funding & 64 & 2,58 \\
            Manufacture of coke and refined petroleum products & 19 & 2,65 \\
            Other manufacturing & 32 & 2,65 \\
            Electricity, gas, steam and air conditioning supply & 35 & 2,72 \\
            Legal and accounting activities & 69 & 2,78 \\
            Activities of head offices; management consultancy activities & 70 & 2,78 \\
            Architectural and engineering activities; technical testing and analysis & 71 & 2,78 \\
            Scientific research and development & 72 & 2,78 \\
            Advertising and market research & 73 & 2,78 \\
            Other professional, scientific and technical activities & 74 & 2,78 \\
            Veterinary activities & 75 & 2,78 \\
            Rental and leasing activities & 77 & 2,78 \\
            Employment activities & 78 & 2,78 \\
            Travel agency, tour operator and other reservation service and related activities & 79 & 2,78 \\
            Security and investigation activities & 80 & 2,78 \\
            Services to buildings and landscape activities & 81 & 2,78 \\
            Office administrative, office support and other business support activities & 82 & 2,78 \\
            Activities of extraterritorial organisations and bodies & 99 & 2,80 \\
            Land transport and transport via pipelines & 49 & 2,86 \\
            Water transport & 50 & 2,86 \\
            Warehousing and support activities for transportation & 52 & 2,86 \\
            Sewerage & 37 & 3,11 \\
            Manufacture of paper and paper products & 17 & 3,15 \\
            Printing and reproduction of recorded media & 18 & 3,15 \\
            Manufacture of other non-metallic mineral products & 23 & 3,31 \\
            Manufacture of rubber and plastic products & 22 & 3,32 \\
            Manufacture of basic metals & 24 & 3,53 \\
            Manufacture of fabricated metal products, except machinery and equipment & 25 & 3,53 \\
            Manufacture of chemicals and chemical products & 20 & 3,60 \\
            Mining of coal and lignite & 5 & 3,84 \\
            Extraction of crude petroleum and natural gas & 6 & 3,84 \\
            Mining of metal ores & 7 & 3,84 \\
            Other mining and quarrying & 8 & 3,84 \\
            Mining support service activities & 9 & 3,84 \\
            \bottomrule
        \end{tabular}
    }
\end{table}

\clearpage

\clearpage

\bibliography{references} 

\end{document}